\documentclass[aps,pra,twocolumn,showpacs,floatfix]{revtex4}
\usepackage{graphicx}
\usepackage{epsfig}
\begin{document}
\setlength{\unitlength}{1mm}
\bibliographystyle{unsrt} 
\title{
Measurement of the parity violating 6S-7S transition amplitude in cesium achieved\\within  $2 \times10^{-13}$ 
atomic-unit accuracy by stimulated-emission detection  }
 \author{J. Gu\'ena, M. Lintz and M.A. Bouchiat }
\affiliation{ Laboratoire Kastler Brossel and F\'ed\'eration de Recherche \\D\'epartement de Physique de l'Ecole
Normale Sup\'erieure,\\ 24 Rue Lhomond, F-75231  Paris  Cedex 05, France } 
 \date{December 9, 2004}
 \begin{abstract}
We exploit the process of
asymmetry amplification by stimulated emission which provides an original method for parity violation (PV)  
measurements in a highly forbidden atomic transition. The method involves measurements of a {\it chiral}, transient, optical
gain of a cesium vapor on the $7S-6P_{3/2}$ transition, probed after it is excited by an intense, linearly polarized, collinear laser,
tuned to resonance for one hyperfine line of the forbidden 6S-7S transition in a longitudinal electric field. We report here a 3.5
fold increase, of  the one-second-measurement sensitivity, and subsequent reduction by a factor of 3.5 of the statistical
accuracy compared with our previous result [J. Gu\'ena {\it et
al.}, Phys. Rev. Lett. {\bf 90},  143001 (2003)]. Decisive improvements to the set-up include an increased 
repetition rate, better extinction of the probe beam at the end of the probe pulse and, for the first time to our knowledge, the
following: a polarization-tilt magnifier, quasi-suppression of beam reflections at the cell windows, and a Cs cell
with electrically conductive windows. We also present real-time tests of systematic effects, consistency
checks on the data, as well as a 1$\%$ accurate measurement of the electric field seen by the atoms, from atomic signals.
PV measurements performed in seven different vapor cells agree within the statistical error.  Our present result is compatible with
the more precise Boulder result within our present relative statistical accuracy of 2.6$\%$, corresponding to  a
$2\times10^{-13}$ atomic-unit uncertainty in $E_1^{pv}$.  Theoretical motivations for further measurements are emphasized and we give a
brief overview of a recent proposal that would allow the uncertainty to be reduced to the 0.1$\%$ level by creating conditions
where asymmetry amplification is much greater.

\pacs  {32.80.Ys, 11.30.Er, 33.55.Be, 42.50.Gy} 
\end{abstract}
\maketitle
%%%%%%%%%%%%
\section{Introduction:  Goal of the experiment}
%%%%%%%%%%%%
Parity Violation (PV) in stable atoms is a manifestation of the weak interaction involving the exchange of a neutral vector boson
$Z^0$  between the electron and the nucleus. It shows up in high precision measurements testing the symmetry
properties of the process of optical absorption, hence in conditions very different from those of high energy experiments
\cite{bou97,bud98}. However, the effects in stable atoms are so small that their detection requires the choice of very
peculiar conditions: {\it i.e.} the use of a heavy atom because of the
$Z^3$ enhancement factor, and a highly forbidden transition to avoid the
electromagnetic interaction completely overwhelming the weak interaction. This explains why we first selected the
highly forbidden $6S\rightarrow 7S $ transition of atomic cesium \cite{bou74} at 539~nm. In the s-orbitals
the valence electron penetrating close to the nucleus, just where the short-range weak interaction can be
felt, is accelerated in the Coulomb potential associated with the nuclear charge Ze, with a strength reinforced by relativistic
effects. Therefore, electron-nucleus momentum transfers of 1 MeV/c can occur, even though the atoms are irradiated by
photons of only 2.3 eV.

In absence of any applied electric field, the 6S-7S transition electric-dipole amplitude is strictly forbidden by the laws
of electromagnetism. We measure the contribution 
which arises from the weak interaction, $E_1^{pv}$. Its order of magnitude is $0.8 \times 10^{-11}$ in atomic units, $ea_0$,
instead of $\sim 1$ for usual allowed transitions in atoms. We compare it with the 6S-7S transition electric-dipole amplitude
induced by an applied electric field, $\beta E$.  These measurements can be used to extract the weak charge
$Q_W^{exp}$  of the cesium nucleus {\it via} an atomic physics calculation, for comparison with the theoretical prediction $Q_W^{th}$ of
the Standard Model (SM) of electroweak unification theory. Thanks to the relative simplicity of the atomic structure of cesium
having a single valence electron,  this calculation is reliable and, owing to recent progress, its
accuracy has now reached 0.5$\%$
\cite{der01,mil01,fla02,fla04}. Moreover, if the measurements are performed on two different hyperfine components, the
results can provide a determination of the nuclear anapole moment  
\cite{bou97,fla80}. 

The first measurements of  $E_1^{pv}$ in cesium were performed by our own group \cite{bou82}. They were followed by
calibrated
\cite{bou88}, more precise ones ($0.5\%$), achieved by the Boulder group \cite{woo97,ben99}.  Today, the latter imply no
significant deviation of
$Q_W^{exp}$ with respect to the SM prediction
\cite{fla04}. By contrast, the reported value of the nuclear anapole moment presents serious discrepancies compared with other
manifestations of parity violating nuclear forces \cite{fla04,CB91}. 
Our goal is to achieve an
independent measurement, as precise as possible, by a different method in order to cross-check the Boulder result.  Our new approach is
based on a pump-probe experiment using two collinear laser beams which operate in pulsed mode, for detection of the forbidden
transition by stimulated emission. Except for the choice of the transition and the use of an applied electric field, this very
different method has nothing in common with the previous PV measurements using fluorescence detection
\cite{bou82,woo97}. We validated this new approach with a 9$\%$ precision measurement in 2002. This was published in
\cite{bou02} where we suggested further improvements of the signal over noise ratio (SNR) by using higher quality cell
windows allowing better cancellation of the reflected beams \cite{jah00} as well as a polarization magnifier
\cite{cha97}. These and additional modifications of the set-up together with a tighter control of the systematic effects, have
resulted in a much better SNR. This paper reports on their implementation and on the subsequent measurement of $E_1^{pv}$
that has now reached a precision of 2.6$\%$. It is organized as follows. In Sect. II, we describe the principles and implementation
of the experiment. The recent decisive improvements of the experimental set-up, and their effect on the SNR are presented in Sect.
III. Thereafter we describe the control and estimation of the systematic effects (Sect. IV). After a summary of the PV data
acquisition and processing, we indicate how we have improved the measurement of the electric field seen by the atoms {\it in
situ} and we present our current PV result together with consistency tests on the data (Sect. V). Finally, the implications of this
kind of measurement are discussed  and we conclude by considering short and longer term prospects (Sect. VI).       
%%%%%%%%%%%%%%
 \section{Experiment: Principles and Implementation}
%%%%%%%%%%%%%%%
\subsection{Detection of a chiral optical gain by stimulated emission  }
 An intense laser pulse, tuned to resonance for one hyperfine component of the $6S_{1/2}-7S_{1/2}$ transition
creates, in a time interval short compared with the $7S$ lifetime (48~ns), a large population difference between the $7S$ and the
$6P_{3/2}$ states, which is immediately detected by a probe laser pulse tuned to resonance with one hyperfine
component of the
$7S_{1/2}-6P_{3/2} $ transition. This pulse is amplified by stimulated emission and compared with a  reference pulse
sent through the vapor once no more atoms remain in the $7S$ and $6P$ states. Figure 1 shows the timing of the
experiment. In this way it is possible to extract the atomic contribution to the amplified probe intensity transmitted
through the vapor. It is the polarization modification of the amplified probe beam which can exhibit Parity
Violation. 
\begin{figure}
\vspace{1cm}
\includegraphics*[width=8.5cm, keepaspectratio =true]{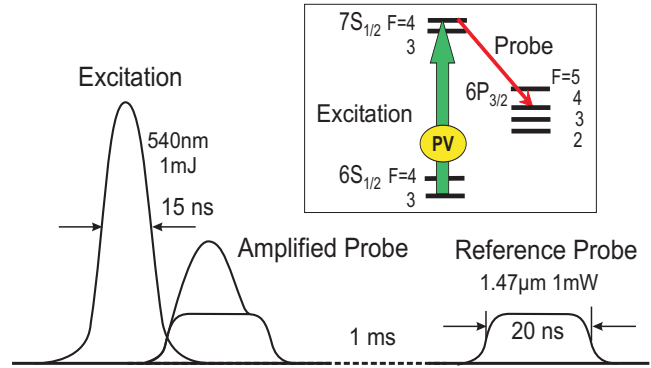}
\caption{ Timing of the experiment repeated for each excitation pulse (150 s$^{-1}$). 
Insert: $^{133}$Cs levels involved (I=7/2, hyperfine splittings: 9.2 and 2.2 GHz 
in $6S_{1/2}$ and $7S_{1/2}$ resp., 251, 201, 151 MHz in $6P_{3/2}$).} 
\end{figure}

 An electric field $\vec E_l$ is applied collinear with the excitation beam. This induces an electric-dipole, ``Stark''
transition amplitude
$E_1^{Stark} = \beta E_l$, large enough for the field to control conveniently the transition rate. The direction of 
$\vec E_l $ and the linear excitation polarization $\hat
\epsilon_{ex}$ define two planes of symmetry of the experiment. Without any PV effect, the eigenaxes of the
excited vapor would lie inside those planes of symmetry. PV  actually shows up as {\it an angular tilt of the
eigenaxes outside of those planes}. The tilt angle, $\theta^{pv}$ equal to the ratio ${\rm Im} E_1^{pv}/E_1^{Stark}$,
odd under reversal of the electric field, is the relevant physical quantity to be measured, the amplitude $E_1^{Stark}$ being
determined independently (see Sect. V. D). From the tilt of the optical axes, it results that the two mirror-image configurations of
Fig. 2, defined by
$\vec E_l , \hat \epsilon_{ex} $ and the linear probe polarization $\hat \epsilon_{pr}$ oriented at $\pm 45^{\circ}$, 
are not physically equivalent: they lead to different amplification factors, {\it i.e.} to a linear dichroism  on the probe transition.
This PV linear dichroism is associated with the pseudoscalar $(\hat
\epsilon_{ex} \cdot \hat \epsilon_{pr})(\hat \epsilon_{ex} \wedge \hat \epsilon_{pr} \cdot \vec E_l)$, {\it i.e.}
{\it chiral}, contribution to the gain of the excited vapor \cite{bou85}. 

At cell entrance the polarizations $\hat \epsilon_{ex}$ and $\hat \epsilon_{pr}$ are alternatively chosen parallel or orthogonal.
Those so-called ``para'' and ``ortho'' configurations both correspond to a linear superposition of the left and right configurations
leading to different optical gains (see Fig. 2).  The tilt angle,
$\theta^{pv} \simeq 10^{-6}$~rad for $E_l =
$1.6~kV/cm, is measured by using a two-channel polarimeter operating in balanced-mode.  The  probe beam is separated by a
polarizing beam splitter cube into two beams polarized at
$\pm 45^{\circ}$ of the incident polarization. The gains of the two photodetectors are adjusted in absence of the
excitation beam  to obtain a null signal difference, hence cancellation of the reference imbalance
$\lbrack (S_1 - S_2)/(S_1 + S_2)\rbrack_{ref} \equiv D^{ref}$ at a level of $10^{-3}$ \cite{gue95}, with a stability
ensuring absence of noise coming from compensation drifts.  When the excitation pulse is switched on with
$\hat \epsilon_{ex}$ parallel (or orthogonal) to $\hat \epsilon_{pr}$, we expect the PV effect to give rise to a polarimeter imbalance
$D^{amp}$, odd under reversal of the electric field \cite{gue97}, since each channel measures the amplified probe intensity in either
one of the two mirror configurations of Fig. 2. For each excitation pulse, the difference $D_{at} = D^{amp} - D^{ref}$
provides  a direct measurement of the PV left-right asymmetry $A_{LR} \equiv D_{at}$, proportional to $\theta^{pv}$ within
a proportionality factor $K$. It is useful to have an explicit form of this factor:
\begin{equation}
A_{LR} = K \theta^{pv} = 2 \theta^{pv} \lbrack\exp{(\eta {\cal{A})}-1)}\rbrack \,,   
\end{equation}
%A_{LR}/2 \lbrack\exp{(\eta {\cal{A})}-1)}\rbrack,   
 with  $\eta$ = 11/12 and $ -11/34$ for the  $7S_{1/2,F=4} \rightarrow
6P_{3/2,F=4}$ transition in the ortho- and para-configurations respectively, and ${\cal{A}}$ is the optical density
for the probe. Even though this expression is valid only in first approximation (see Sect V D and \cite{bou96}), it
describes well all the observed features of the amplification process. Moreover, the factor  
$K$ is eliminated in the calibration procedure allowing us to convert the {\it imbalance} into an absolute
$\theta^{pv}$ {\it angle}. We achieve this by performing with a Faraday rotator an angular tilt of the
excitation polarization with respect to $\hat \epsilon_{pr}$, of a precisely known magnitude $\theta_{cal}$, and by
measuring the resulting imbalance: $K \theta_{cal}$.
\begin{figure}
\includegraphics[width=60mm, keepaspectratio=true]{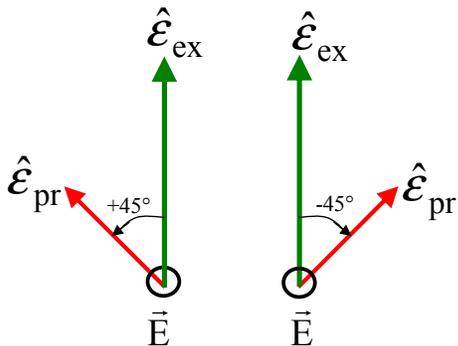} 
\caption{ Illustration of two mirror-image configurations (left and right) that exhibit a gain asymmetry due to 
parity violation. The
experiment is performed with either $\hat \epsilon_{ex} \parallel$ or $ \perp $ to $\hat \epsilon_{pr}$ so as to 
yield directly the gain asymmetry (see text).   }
\end{figure}

\subsection{Important differences with respect to the Boulder experiment}
 Conceptually the present experiment \cite{gue97,gue98,bou02} is very different from  
that of the Boulder group \cite{woo97,woo99}. This is illustrated by a set of features summarized
below:  
\begin{itemize}
\item We have a direct measurement of a left-right (LR) asymmetry, $A_{LR}$, instead of a transition rate 
superimposed on a background of 25$\%$.
\item As shown by Eq. 1 we benefit from an amplification of the (LR) asymmetry with the optical density for the probe
${\cal{A}}$ (see Eq. 1 and\cite{bou96}). As a result the parameter we measure $A_{LR}$ is an increasing function of $E_l$
instead of one  falling as $1/E_l$.
\item We employ a longitudinal $\vec E_l$-field configuration instead of a transverse $\vec E$ and $\vec B$-field
configuration.  The absence of Stark-$M_1$ interference in this 
 configuration suppresses a potential source of systematic effect troublesome in a
transverse E-field.  
\item Our experiment is not limited to the two $\Delta F = \pm 1 $ hyperfine components but can be extended to all of them.
\item Our experiment is absent from line-shape distortion  and line-overlap problems. 
\item Our line-shape independent calibration is performed continuously: the PV signal (PV alignment) is calibrated
in real time by a signal of the same nature (Stark alignment \cite{bou03}), as opposed to the calibration of the Boulder
experiment, made difficult by incomplete line-resolution, saturation and background effects \cite{woo99}.
\end{itemize}                     
  Direct detection of an angular anisotropy in the $7S$ state, which ensures specificity of the signal, is obtained
to the detriment of the Signal-to-Noise ratio (SNR); the latter is definitely lower than that achieved in Boulder, 
$\sim 1$ in 10~s \cite{woo97}. However, with the SNR improvement reported in this paper, the sensitivity of our
pump-probe method already appears adequate for APV measurements at the 1$\%$ level and we mention (Sect. VI)
an extension of the method expected to improve considerably the quantum, shot-noise limit. 
  
\begin{table*}
\caption{Criteria and parameter reversals defining the complete PV signature, with a binary variable $\sigma_i$
attached to each reversal. One elementary state of  the experiment, with no parameter reversal,   
lasts 30 laser shots. The period of each reversal is indicated for a
repetition rate of the excitation laser of 150 Hz. The sequence $(\pm \theta_{cal}, \pm E_l)$ is repeated four times 
before we perform reversal 4, and the sequence up to reversal 6 is repeated five times. Lower box: the eight different
polarization configurations ($\hat
\epsilon_{ex},
\hat \epsilon_{pr}$) used for the measurements. We indicate the polarization orientations of the excitation (large
arrow) and probe (small arrow) beams.    }
\begin{center}
\begin{tabular}{|p{4cm}|p{2cm} | p{4cm}|p{3.5cm}|p{1.8cm}|c|} 
\hline Criteria, reversal & PV signature& Selection of & against& Nb of exc. shots & period \\ \hline

$1) \hspace{4mm} D^{amp}=\lbrack \frac{S_{1}-S_{2}}{S_{1}+S_{2}}\rbrack^{amp}$& & ~ Polarimeter imbalance  & 
${\rm Intensity,~population}$ ~~~~~~~~~~~~~~~~~~~~ & 1 & \\
\hline

$2) \hspace{4mm} D_{at}=D^{amp}-D^{ref} $ & &Imbalance of atomic origin $\equiv $ LR asymmetry & Non atomic & 1  &1
ms\\
\hline

$ \hspace{6mm} \pm \theta^{cal}  ~~~~~~(\sigma_{cal}= \pm 1)$ & ~~~~~Even & ~Calibration imbalance 
& & 2x30 & 0.4~s\\
& & & & &\\ \hline

$3) \hspace{2mm}\pm E_{l} ~~~~~~~~~ (\sigma_{E}= \pm 1)$ & ~~~~~Odd & ~ $E_{l}$ Odd PV effect & Most PC effects
& 2x 60 & 0.8 s\\
& & &(Stark-Stark) & &  \\\hline

$4) \hspace{1mm}\pm$tilt $\hat{\epsilon}_{pr}^{out}~$& ~~~~ Odd & True polarization effect &
Instrumental defects  & 2x 4x120 & 7 s \\
$~~~~~~~~~~~~~~~~~~~~~~(\sigma_{det}= \pm 1)$ & & &~~(EMI, geometrical) & &  \\
\hline

$5) \hspace{0mm} \hat{\epsilon}_{pr}^{in}\parallel,\perp\hat{\epsilon}_{exc}$  & ~~~~ Even
&Linear dichroism of atomic    & Optical rotation & 4x 4x120 & 14.5 s\\ 
$~~~~~~~~~~~~~~~~~~~~~~(\sigma_{pr}= \pm 1)$& & ~~~origin (alignment) &~~~ (e.g. Faraday) & & \\\hline

6) \hspace{2mm}   $\hat{\epsilon}_{exc} {\lbrace x,y} \rbrace  $ &~~~~Isotropic & ~Rotational invariant & Stray transverse
  & 4x5x  & 5 min \\ 
 7) \hspace{2mm} $\hat{\epsilon}_{exc}
{\lbrace u,v} \rbrace$  & & &  $\vec B_{\perp} \& \vec E_{\perp}$ fields& (4x4x120) & \\  \hline
$8) \hspace{3mm} \pm ~ {\rm  tilt~of~cell~axis}$  & ~~~~~Even&~~Incident excitation &  Back-reflected  &$~\sim 10^6$ &
90 min\\
  $ ~~~~~~~~~~~~~~~~~~~~~(\sigma_{\psi}=\pm 1)$ &  &~ beam& excitation  beam & & \\\hline 

\end{tabular}
  
\begin{tabular}{|p{6.0cm} |c|c|c|c|}
\hline
\hspace{0mm}4 ($\hat{\epsilon}_{exc},\hat{\epsilon}_{pr}$) $\parallel \& \perp$ configurations & 
\scalebox{1}[3]{$\Uparrow$}
$\uparrow$
\hspace{2mm} \& \hspace{2mm} \scalebox{1}[3]{$\Uparrow$} $\rightarrow$ &
\rotatebox{90}{\scalebox{1}[3]{$\Uparrow$} $\uparrow$}   \&  \rotatebox{90}{\scalebox{1}[3]{$\Uparrow$} $\rightarrow$}
&
\rotatebox{45}{\scalebox{1}[3]{$\Uparrow$} $\uparrow$}   \&  \rotatebox{45}{\scalebox{1}[3]{$\Uparrow$} $\rightarrow$}&
\rotatebox{-45}{\scalebox{1}[3]{$\Uparrow$} $\uparrow$}   \&  \rotatebox{-45}{\scalebox{1}[3]{$\Uparrow$} $\rightarrow$}
\\ \hline

\end{tabular}
\end{center}
\end{table*}

\subsection{Experimental methods and set-up}
Table~I summarizes all the parameter reversals which allow us to identify the PV signal owing to its well defined
signature. They fall into three categories: i) the two most rapid ones, 1 and 2, allow us to isolate the LR asymmetry of
purely atomic origin,  $A_{LR} \equiv D_{at} =  D^{amp} - D^{ref}$ via a difference performed at the millisecond
time scale;  ii) the next ones, 3 to 5, select the
$E_l$-odd, polarimeter imbalance behaving like a linear dichroism of the atomic vapor with axes at $\pm
45^{\circ}$ of the probe polarization; iii) then, 6 and 7, exploit the  invariance of the PV effect under simultaneous
rotation of the excitation and probe polarizations about the common beam direction
\cite{bou03}. Thus, the measurements are performed in eight different configurations of $\hat \epsilon_{ex}\,,\hat
\epsilon_{pr}$ represented in Table 1 lower box.  An additional fast reversal, not acting on the PV signal, is the    
$\hat \epsilon_{ex}$ tilt by an angle $\theta_{cal}$, of 1.76~mrad, required for calibrating the polarimeter imbalances. When the
laser repetition rate is adjusted to 150~Hz, total completion of all these reversals takes 5 minutes and provides two
``isotropic'' determinations of $\theta^{pv}$. Finally, after data acquisition for a period of typically 90 min, we
reverse the slight tilt ($\sim
$3 mrad) of the cell axis with respect to that of the laser beams (see Sect. II D below ). Hence our experimental
procedure provides an 8-fold signature for the PV signal. More details on data acquisition and processing are given in
Sect. V.

Figure 3 shows a general schematic view of the experiment. A detailed description of the various elements was 
presented in 1998 
\cite{gue98}. Since then, the important modifications we have made mainly concern
the cesium vapor cells and the electric field generator. In particular, previously the cells were made of glass, with 9
electrical feedthroughs and internal electrodes. They are now made of a simple alumina or sapphire tube (length:
83~mm, internal diameter 10~mm) with a sapphire window glued at each end \cite{sar89} and a side arm containing
Cs metal. Surface conductivity of the cesiated walls is considerably lower than with cesiated glass
\cite{bou99}, which allows us to use external electrodes \cite{jah01}. 
Problems encountered with glass cells in our operating conditions (Cs density $\sim 10^{14}$ at/cm$^3$, excitation energy 1.8 mJ, $E_l
\simeq 1.7$~kV/cm) were solved by such cells: i) surface conduction currents, ii) photoionization of Cs$_2$ dimers
\cite{bou92} followed by charge separation in the $\vec E_l$ field, iii) loss of transparency of the windows. The calculated
field map in the external-electrode set-up, together with the production of the flat-top reversible high voltage pulses
\cite{notpul}, are shown in ref. 
\cite{jah01}.     
  
However, the first signals in the sapphire cells showed the presence of unacceptable stray fields resulting from
photoionization at the windows during the excitation pulse. 
We succeeded in reducing them by grooving circular rings on the internal surface of the wall: these prevented 
charge multiplication of  the longitudinally accelerated photoelectrons 
striking the wall at grazing incidence. After this decisive step \cite{gue02}, significant PV measurements could start and a preliminary
$9\%$ accurate measurement validating our detection method \cite{bou02} was achieved.

We, now, present the further decisive improvements brought to the set-up since these initial measurements. 
\begin{figure*}
\centerline{\epsfxsize=170mm \epsfbox{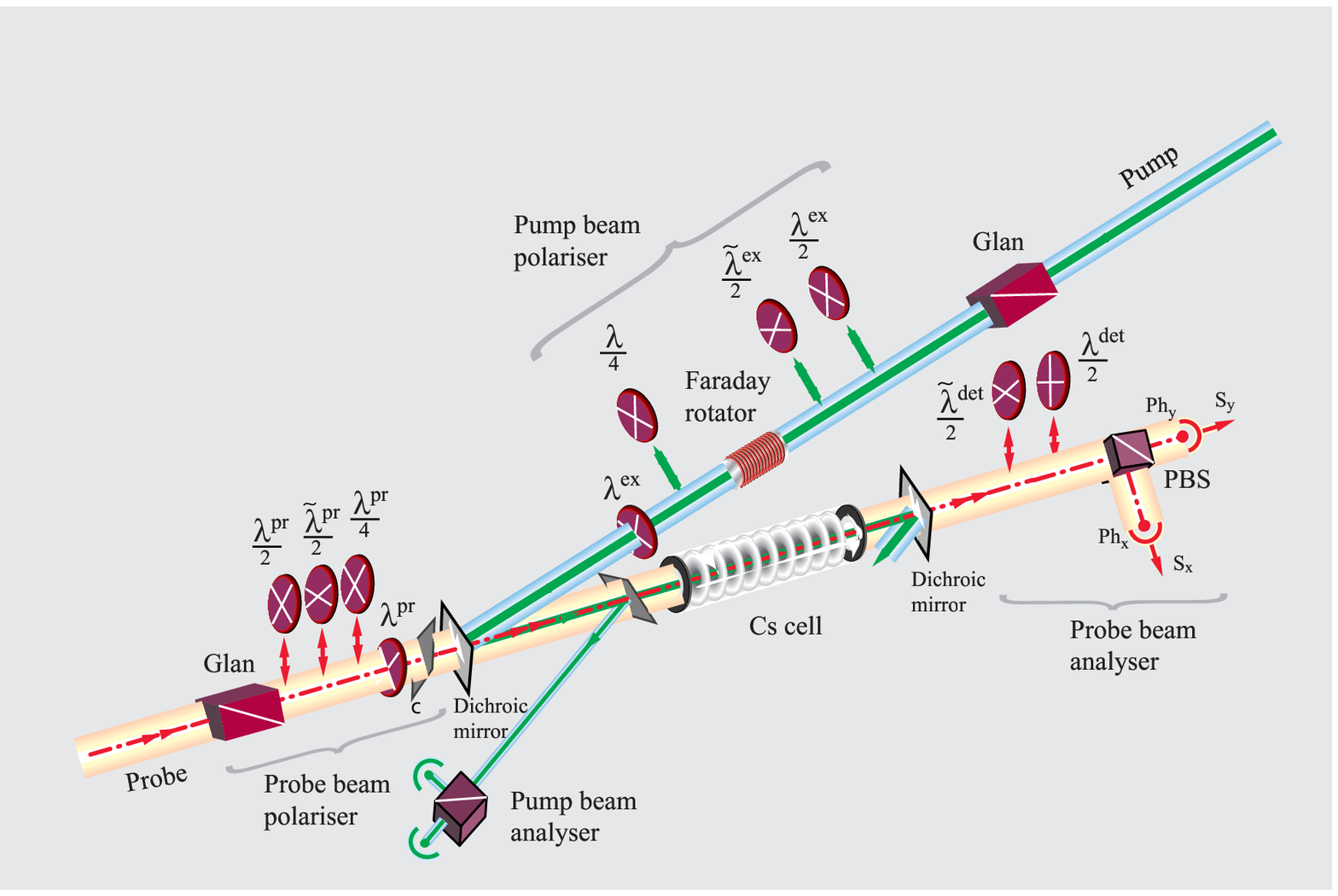}}
\caption{ Schematic view of the experiment. $\lambda /2, \tilde{\lambda} /2, \lambda /4 $: 
insertable half- and quarter- wave plates for controlling the pump and probe polarizations (superscript ex and pr
resp.). $\lambda /2$ and
$\tilde \lambda /2$ allows for reversals 6 and 7 in Table 1, respectively. The $\lambda /4$ are used when circular
polarization is required for initial frequency tuning of the pump beam. $\lambda$: tilted wave-plates for
birefringence compensation. $\lambda^{det}/2$: performs reversal 4 in Table 1. $\tilde{\lambda}^{det}$/2 :
restores the probe when input $\tilde{\lambda}^{pr} /2$ is inserted. PBS: polarizing beam splitter cube with axes
at $\pm 45^\circ$ of axes x,y (polarimeter analyzer). The pump beam analyzer performs analogous role for the
pump beam.
$Ph_x, Ph_y$: InGaAs photodiodes, providing the polarimeter signals $S_x, S_y$ resp. ($S_1, S_2$ in text). Dichroic mirror: dielectric
mirrors reflecting the pump ($R > 99.9\%$) while transmitting
$\sim 95\%$ of the probe. Glan: Glan-air polarizer. C: glass plate to compensate linear dichroism.
Cs cell: cell with external ring electrodes. }
\end{figure*}
%%%%%%%%%%%%%%        
\section{Recent experimental progress}
%%%%%%%%%%%%%%

\subsection{Improvement of the excitation laser source}
\subsubsection{Limitations on the frequency stability coming from the reference cavity}
Since our calibration procedure eliminates any dependence of the signals on the line shape, slow drifts of the
excitation and probe frequencies are not a source of systematic error. However, it is important that the lasers stay at resonance to
preserve the optimum of the SNR, particularly in a PV experiment where the data acquisition lasts over long periods of time.

The probe beam frequency is stabilized on a hyperfine component of the $7S-6P_{3/2}$ transition, using polarization spectroscopy in
an auxiliary Cs cell in which a discharge continuously populates the $6P_{3/2}$ level. An analogous method is not possible on
the forbidden $6S-7S$ transition. The excitation frequency is stabilized on an external Fabry-Perot Cavity (FPC). The FPC
is tuned at resonance for the  $6S_{F=3}\rightarrow 7S_{F=4}$ transition using dispersion-shaped signals
provided by the pump-probe PV set-up itself, either optical rotation, by
temporarily making the excitation beam circularly polarized (case of the  $6S_{F=3}\rightarrow 7S_{F=4}\rightarrow
6P_{3/2,F=5}$ transition) or birefringence, by temporarily making the probe beam circularly polarized (case of the 
$6S_{F=3}\rightarrow 7S_{F=4}\rightarrow 6P_{3/2,F=4}$ transition). This allows us to tune initially the excitation
frequency and then control it at regular time intervals. Although the vessel containing the reference  FPC is
evacuated and temperature stabilized, the excitation laser frequency drifts by typically a few  megahertz per minute.
We first used a correction procedure assuming a linear drift, but the slope was not constant enough for the
approach to be reliable. To do better we stabilized the reference cavity  on an iodine molecular line.   

\subsubsection{Long term frequency stabilization of the cavity on a $^{127}I_2$ line}
 By observing the fluorescence of Iodine in the
region of interest and by  using the theoretical spectrum of $^{127}I_2$ given by the program {\it IodineSpec}
\cite{bod00} supplied by {\it Toptica}, we found that the line the closest to the $6S_{F=3} \rightarrow 7S_{F=4}$ hyperfine
transition was the very weak hyperfine component (a15) belonging to the $J^{''} = 36, \nu^{''}=1 \leftarrow J^{'} = 37,
\nu^{'}= 31$ rovibrational transition (37 P (31-1)), the actual frequency difference being $\Delta \nu \simeq - 300   $~MHz.  The
method of saturation spectroscopy is required to resolve the hyperfine lines of $I_2$. A long iodine cell (50 cm) is necessary to
increase the absorption signal and saturation effects are enhanced by focusing the laser beams into the cell with 30~cm focal
lenses.
 
A fraction of light (13 mW) is taken from the  539~nm cw dye laser (before its pulsed amplification, see Sect. 2 below) and
enters the set-up through a polarizing beam splitter cube, as schematized in Fig. 4. A large fraction of this beam, {\it i.e.} the
pump, is passed twice through an acousto-optic modulator (AOM1), shifting the frequency by
$\Delta
\nu_1 = - 2\times 200 $~MHz. The small fraction of it, (the probe) is passed through a
second acousto-optic modulator, AOM2, shifted in frequency by $ 
\Delta \nu_2 \sim - 200 $~MHz  and superposed with the pump  in  the $I_2$ vapor along a counterpropagating
path. The saturation spectroscopy signal appears  when the sum of the laser frequency $\nu_{laser}$ and
$\frac{1}{2} (\Delta \nu_1 +\Delta \nu_2)$ is resonant with the iodine vapor. To achieve enough
sensitivity, a frequency modulation (amplitude 2.0 MHz, modulation
frequency 19 kHz) is superimposed on $\Delta \nu_1$. Then, by lock-in detection of the transmitted probe
intensity at the pump modulation frequency, we can observe the derivative of the absorption signal. This symmetric
dispersion-shaped signal  (Fig.5) is used as an error signal in a feedback loop ensuring long term stabilization of the external 
FPC to the center of the hyperfine $6S_{F=3} \rightarrow 7S_{F=4}$ Cs transition after initial $\Delta \nu_2$ adjustment. 
In this way, 
frequency drifts are suppressed (i.e. smaller than the cw laser spectral width, $\sim$ 1~MHz) without  interruption
of the PV data acquisition, over periods as long as several hours.  Note that the use of two AOMs is very helpful,
since it ensures excellent rejection of any Doppler background and absence of stray light modulated at the signal
frequency without requiring the use of a reference beam, most often sujected to drifts.
\begin{figure}
\begin{center}
\includegraphics*[height=8cm, width=8.7cm, keepaspectratio=true]{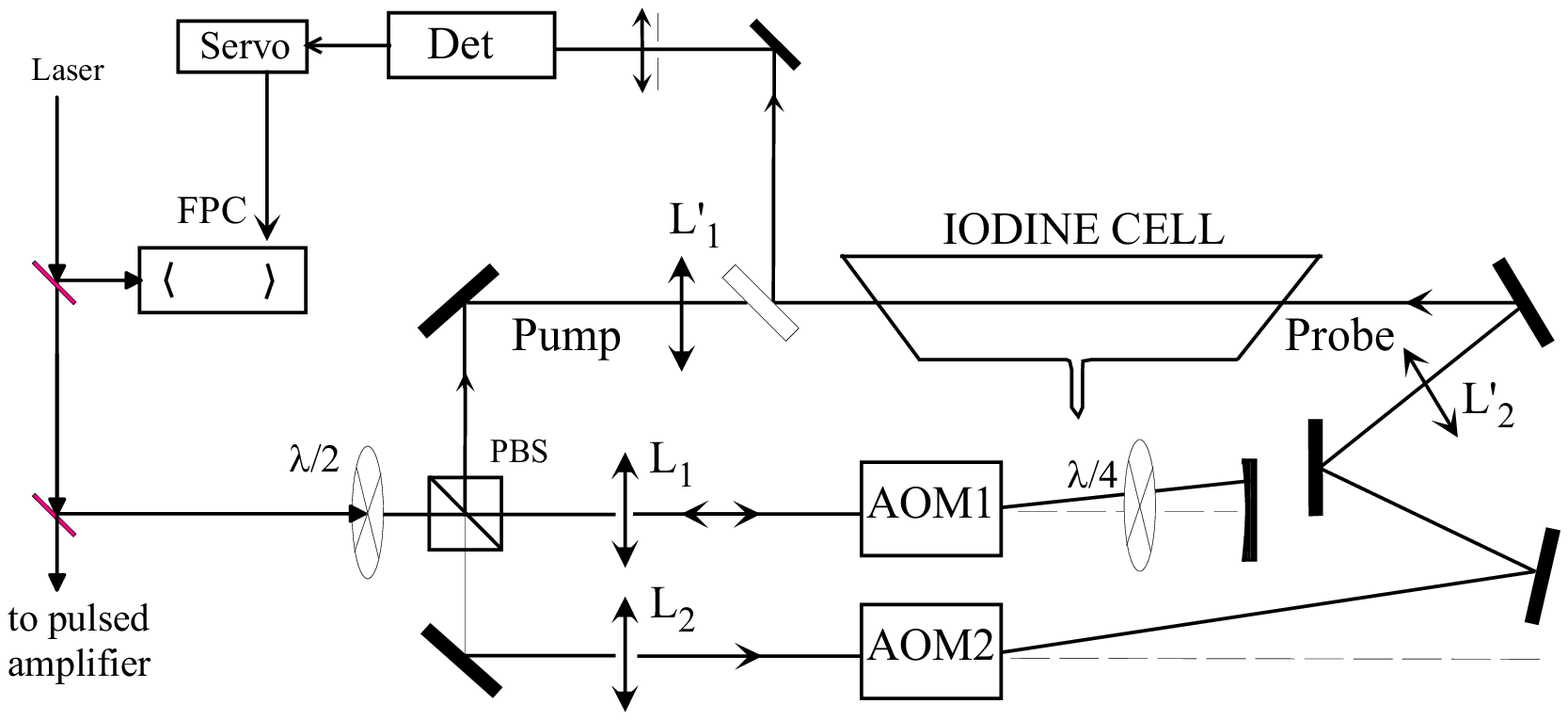}
\caption{ Experimental set-up for excitation-laser frequency stabilization by saturation spectroscopy of $I_2$. 
FPC: reference Fabry Perot cavity. PBS: Polarizing beam splitter cube. AOM 1, 2: acoustooptical modulators.}
 \end{center}
\end{figure}
\begin{figure}
\includegraphics*[height=6cm, width=6cm, keepaspectratio=true]{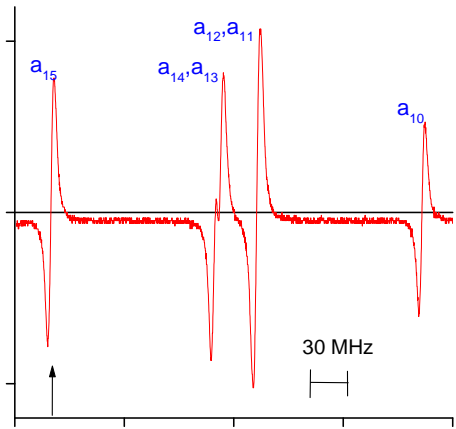}
%\centerline{\epsfxsize=60mm \epsfbox{Spectre-Iode.eps}}
\vspace{5mm}
\caption{ Iodine saturation-spectroscopy signal used to stabilize the reference Fabry Perot cavity,
obtained with the set-up shown in Fig. 4. The arrow indicates the hyperfine transition 37 P (31-1) of $^{127}I_2$ 
used for locking on the $6S_{F=3} \rightarrow 7S_{F=4} $ transition.} 
\end{figure}

At the beginning of a PV data acquisition, the shift $\Delta \nu_2$ is adjusted so as to maximize the probe
amplification in the cesium cell or, for higher sensitivity, to cancel out the relevant dispersion shaped signal. 
Because of the a.c. Stark shift $\Delta \nu(i_{ex})$ induced by the excitation laser, the value of the adjusted shift
$\Delta \nu_2$ depends on the excitation energy $i_{ex}$:  $\Delta \nu_2(i_{ex}) =  - 2\Delta \nu(i_{ex})$, since
$\Delta \nu_1$ is kept fixed. The a.c. Stark shift is shown in Fig. 6.  Since the excitation and the
probe pulses overlap, the intense excitation pulse shifts both the excitation and the probe transition frequencies,
by  $\Delta \nu_{ex}$ and $\Delta
\nu_{pr}$ respectively.  In our operating conditions, $\Delta \nu_2$  is thus a linear combination of both shifts. 
Whatever the value of $i_{ex}$, the probe laser frequency remains locked to the atomic transition observed in a
reference cesium cell where the excitation beam is absent.  Consequently, in
presence of the excitation beam, the probe beam is no longer resonant for the atomic zero-velocity class, but 
becomes resonant for atoms of velocity
$c \Delta
\nu_{pr}(i_{ex})/\nu_{pr}$. The maximum of amplification occurs when the excitation beam is resonant with that same velocity class.
  The laser shift required to be at resonance is  then $\Delta \nu_{laser}(i_{ex}) = \Delta
\nu_{ex}(i_{ex}) -\frac{\nu_{ex}}{\nu_{pr}} \Delta \nu_{pr}(i_{ex})$. Hence, the shift of $\Delta \nu_2$
accompanying the $i_{ex}$  variations is interpreted as:
\begin{equation}
  \Delta \nu_2(i_{ex}) = -2 \Delta \nu_{ex}(i_{ex}) +  2 \frac{\nu_{ex}}{\nu_{pr}} \Delta \nu_{pr}(i_{ex}) \;.             
\end{equation}
Since ${\nu_{ex}}/{\nu_{pr}}  = 2.72 $, we note the large sensitivity of $\Delta \nu_2(i_{ex})$ to the a.c. Stark shift
of the probe transition. 

At first sight the large a.c. Stark-shift  {\it difference} between the two hyperfine lines shown on Fig. 6 might look  
surprising. Actually an {\it additional} a.c. Stark shift, of opposite sign, is induced by the probe
beam when it is amplified during propagation through the vapor, the proportionality to $i_{ex}$ resulting from the
amplification. After a careful study, out of the scope of this paper, we have found that this effect, greater for the 3-4-5 system
than for the 3-4-4 one which has a probe gain twice smaller, explains the overall Stark-shift difference.  
 \begin{figure}
 \includegraphics[width=70mm, keepaspectratio=true]{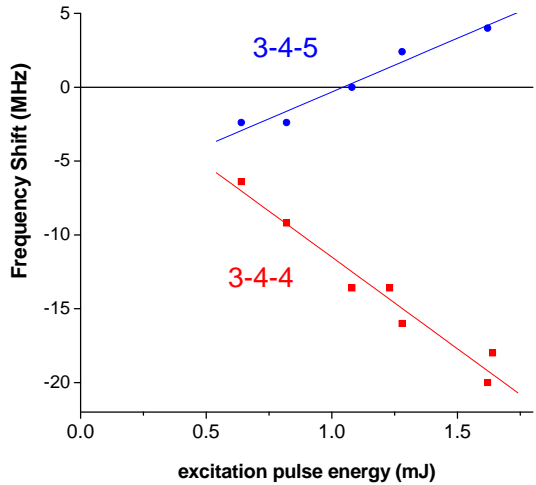} 
\caption{ a.c. Stark shifts $\Delta \nu (i_{ex})$ induced on the $6S_{F=3} \rightarrow 7S_{F=4}
\rightarrow 6P_{3/2,F'}$ resonance {\it {versus}} the excitation energy $i_{ex}$, deduced from the shift of AOM2 in
the iodine set-up of Fig. 4 (see text). Temporal pump-probe overlap: 10~ns. The origin on the vertical axis
is chosen arbitrarily and is different for both transitions.}
\end{figure}
 
\subsubsection{Improvement of the spatial profile of the pulsed beam }
The beam delivered by the cw dye laser is pulsed amplified by passing through a commercial system (Lambda Physik FL 2003)
with three rectangular cells through which circulates a dye solution (Coumarin 153 in Methanol) pumped by a XeCl 
excimer laser  (18 ns long pulses, 100 mJ at 308 nm) at a maximum repetition rate of 200 Hz. With transverse
pumping, optical alignments necessary to obtain the desired ideal circular spatial profile of the beam proved to be rather critical.
 We have tried for the last amplifier to use a Bethune cell which ensures pumping with cylindrical symmetry. A nice
circular shape was obtained, but only at repetition rates below 30~Hz. Beyond this rate, an unacceptable jitter of the
beam position attributed to the turbulence of the liquid flowing
through the cell  forced us to abandon. Trying back the rectangular cell with the  dye circulator
dedicated to the Bethune cell, a more powerful one with an adjustable flow rate \cite{note2B}, we could obtain a stable
circular profile at repetition rates up to 180~Hz, with only little dependence on the day-to-day realignments and on
the aging of the dye-solution. Figure 7 illustrates such a typical profile.    

\begin{figure}
\begin{center}
\includegraphics*[width=8.6cm, keepaspectratio=true]{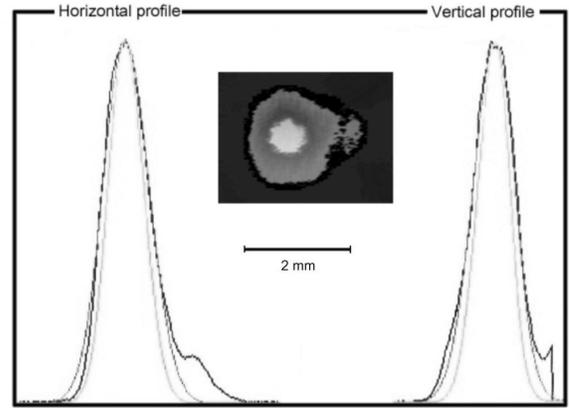}
\vspace{-5mm}
\caption{ Spatial profile of the pump excitation beam. Inset: typical image of the beam transverse-section obtained
with the CCD camera analyzing a fraction of the beam at the input of the Cs cell. Integrated profiles along x and y, horizontal and
vertical axes, respectively. Dark line: measured profile; gray line: gaussian fit with beam waist radius of 0.90 mm; light grey line:
for comparison, probe beam gaussian profile of radius 0.70~mm.}
\end{center}
\end{figure}
  
\subsection{Increasing the probe beam extinction ratio }
\subsubsection{Motivations}
The optical switch which produces 20 ns long optical pulses from the continuous probe laser source at 1.47 $\mu$m, is a
key-component of the set-up.  It turns out that the noise performances of our detection method depends crucially on its extinction ratio.
This integrated device \cite{car88} driven by low voltages, has rapid rise- and fall-times ($\leq 1 $~ns) and its
extinction ratio 
 is usually as low as
$10^{-3}$.  But, probably due to humidity changes, occasionally it rises to several times $10^{-3}$, rendering 
reliable  measurements impossible. Indeed, the leakage photons going through the ``closed'' optical switch 
continue to probe the Cs vapor. They participate in the detected signal for the time characteristic of the
photocurrent integration operated 
 by the dual detection chain at each probe pulse \cite{hri89}.   Since the
$6P_{3/2}$ state, of the 5$~\mu$s effective lifetime due to resonance radiation trapping, is progressively
populated in the early stages of deexcitation of the vapor \cite{note6P}, the time-integrated probe intensity
undergoes a prohibitively large absorption.  The baseline of the charge integrated pulse (20~ns rise time followed by a
50 $\mu$s exponential decay) becomes distorted and fluctuates. This causes noise that is not completely filtered
out by the gaussian pulse shaper (adjusted with $\tau_i= \tau_d = \tau =1~ \mu$s for optimal noise rejection
\cite{hri89}).       
To prevent noise blasts due to the unpredictable behaviour of the optical switch, we
installed a Pockels cell in series with it. 
\begin{figure}
 \centerline{\epsfxsize=65mm \epsfbox{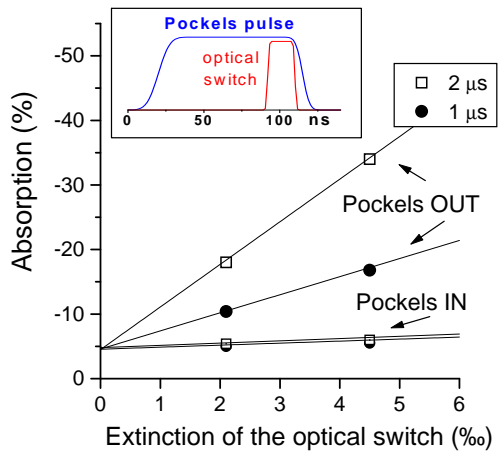} }
\vspace{5mm}    
\caption{\footnotesize   Reduction of the probe-beam absorption thanks to the Pockels cell for two values of the time
constant of the gaussian pulse shaper, $\tau = 1$ and 2 $\mu$s. Inset: timing of the light pulses produced by the Pockels cell and by the
optical switch. } 
\end{figure}
 \begin{figure}
\centerline{\epsfxsize=85mm \epsfbox{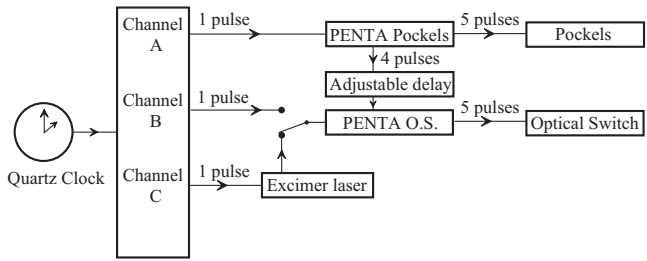}}
\caption{ Scheme of the electronic circuitry feeding five synchronized trigger pulses to drive the
Pockels cell (``Penta Pockels'') and the optical switch (``Penta OS'') at each amplified pulse followed by four reference
pulses. Each channel (A, B, C...) has an adjustable delay. }
\end{figure}
 
\subsubsection{Operating conditions of the Pockels cell and the optical switch}
We placed the Pockels cell (PC) right at the input of the optical switch (OS). This provides the advantage of reducing
considerably the periods  of illumination of the OS, probably a wise protection against possible photo-creation of
color centers in the integrated waveguides. The transmission is 91
$\%$ and the extinction ratio  0.4$\%$. The PC is driven by a fast pulse generator (fall and rise time of about 20 ns, minimum
pulse duration 100 ns
\cite{note3A}). As shown in the inset of Fig. 8,  the beam is let through $\sim $80 ns before the OS opens but it is 
interrupted 
shortly after the OS closes.   
 Figure 8  also illustrates the effect of the PC on the absorption measurements. We plot the measured absorption with
or without the PC versus the extinction ratio of the OS, varied at will by changing the driving voltage. When the PC
is operated, the probe absorption is considerably reduced and the initial dependences on both the extinction ratio of
the OS and the time constant of the shaper disappear. 

 \subsubsection{Trigger system for synchronized operation of the cell and the switch}
The synchronisation of the whole system presented a technical difficulty because the first probe pulse amplified by the vapor is
followed by four reference probe pulses separated in time by about 1 ms, for making negligible the photon shot noise on the reference
measurements. It is important that the shape of the reference pulses be identical to that of the first
probe pulse. This requires excellent synchronization between the pulses which drive the Pockels cell and the optical
switch. On the other  hand, the first probe pulse must be triggered by the excimer laser driving the excitation pulse
so as to avoid a temporal  pump-probe jitter ($\sim 5~$ns shot-to-shot fluctuation of the excimer laser thyratron). 
The triggering system implemented is sketched in Fig.~9. It has two different operating modes, one
requiring operation of the excimer system, the other, which does not, being used for preliminary procedures (probe beam
alignment and polarimeter adjustment).

\subsection{A birefringence-free polarization-tilt magnifier}
\subsubsection{Principle of polarization-tilt magnification}
In an earlier work \cite{cha97}, we suggested using a dichroic component to amplify the tilt angle acquired by the
probe beam after its passage through the excited vapor. The idea
 is simply to transmit differently the polarizations along two orthogonal directions, the direction of the
incoming beam and the orthogonal direction.  In this way we can attenuate the polarization parallel to the incoming
probe polarization, whilst letting through the orthogonal polarization which carries  the important information, {\it
i.e.} the tiny component resulting from the tilt induced by the vapor.

Let us suppose that $\hat x$ is the incoming probe polarization and that the transmission coefficients of our dichroic device for
the amplitude of the field are
$t_x = 0.5$ and $t_y \simeq 1$ in the two eigen-directions $\hat x$ and $\hat y$. It is seen easily that
after passage through this device, the angle between the light polarization and the $\hat x$ direction is multiplied by 2.
The light intensity reaching each detector is divided by 4.  Such a ``polarization magnifier'' does not automatically lead to an
increase of the SNR. In the shot noise limit, we cannot
expect any SNR improvement, except if we increase the intensity of the probe beam at the input of the cell provided that the
asymmetry amplification is not damped under the effect of saturation by the probe beam intensity \cite{cha98}.   
Actually the power of our color center laser can be increased by a maximum factor of $\sim 2.5$. With the 1.5~mW power then 
available in the Cs cell (100~mW at the output of the laser), saturation effects for the probe transition $7S_{F=4} - 6P_{F=4}$ 
the best suited to PV measurements remain small, as expected
\cite{cha98}. If photon shot noise is dominant, the SNR expected improvement is $\sqrt {2.5}$, which was worth to try. 

We note that the magnifier plays the same role as the ``uncrossed polarizer'' in the quasi-extinction method of polarimetry. Thereby,
this dichroic element allows us to combine the advantages of two well-known operating modes of polarimetry:  
quasi-extinction mode and balanced-mode. In the former, intensity noise is reduced by optimizing the
uncrossing angle, while the latter allows one to suppress all common-mode noise.   
\subsubsection{Realization}
  Initially, we considered the possibility of using a single plate with a dichroic multilayer coating \cite{cha97,san04},
but this solution suffered from problems linked with the large, incidence-dependent, birefringence of the coating.
We finally turned to a stack of 4 or 6 silica plates at Brewster incidence. This arrangement yields the required
magnification factor ($\sim $ 2 or 3) and birefringence is 
acceptably small provided that interference between multiple reflections are
suppressed by using wedged, tilted plates, so that no two surfaces are parallel \cite{lin04}.

As explained previously, the polarization of the incoming probe beam at cell entrance determines the polarization
direction that has to be attenuated before reaching the polarimeter. For practical reasons, the device operates in a
fixed position, with the incidence plane vertical. Nevertheless, we can use it  in any of the four possible directions of
the probe polarization,  thanks to two additional half-wave plates which restore the polarization direction before
the magnifier.     
 
 We have found that the device with either four or six plates,   
operates satisfactorily with very stable characteristics.  Without the magnifier, the calibration 
left-right asymmetry for a tilt $\theta_{cal}$ of $\hat \epsilon _{ex}$ with respect to $\hat \epsilon_{pr}$  is 
compatible with the expected value  
$2 \theta_{cal} \times \lbrack\exp{(\eta {\cal A}) -1 )}\rbrack $ (Eq. 1). With the magnifier in position, the measured left-right
asymmetry is amplified with respect to that value by the expected factor
$t_y/t_x$,  for the eight pump-probe configurations.
As to the gain in SNR, we estimate it to be about 1.5 to 2 with the 4-plate device, hence barely larger than what we could expect in
the shot-noise limit. The 6-plate device  improves this result slightly, by at most
$20\%$. This last result is easily understood were shot-noise dominant in our operating conditions: 
i) since we could not increase
the input intensity sufficiently to compensate for the twice higher attenuation factor($ t_x^2 \approx 1/8 $ instead of 1/4), the
small intensity-independent contribution of noise increased accordingly  and  ii) for the larger probe beam intensity required for  
better efficiency of the device, atomic saturation
started to show up on the asymmetry amplification.    

Finally, we mention improvements on the probe beam part of the experiment. Residual etalon effects inside the
polarimeter have been cancelled out. The short-term stability of the servo loop used to stabilize the probe laser frequency has been
improved. Since the detected probe flux is lower when the magnifier is in use, (even though the incoming probe
intensity is higher), we have  increased the gain of the charge-integrating  preamplifiers, by changing the input
capacitor from 7 to 4 nF, so as to use the whole dynamic range of the detection chain.            
\subsection{Control of the beam reflection at the cell windows }
Due to the high refraction index of sapphire (n=1.77), reflection of the input
laser power at each window is about 15$\%$ and can be a source of losses on the 7S excited atom density (reflection
at the entrance window) or of  uncontrolled contributions (reflection at the output window). One might also worry
about interferences taking place {\it between} the two windows which are precisely mounted, normal to the cell axis
(tilt less than 1~mrad).  
\subsubsection{Interferences inside and between the windows: source of noise}
Important progress has been achieved with the realization of the extinction of the excitation beam reflected by
the windows. Depending on its parallelism, any cell window behaves more or less as a temperature-tunable Fabry
Perot etalon 
\cite{jah00}. In the very first cell used for measurements, imperfect parallelism prevented us from obtaining a
reflection coefficient lower  than $\simeq 5 \times 10^{-2}$.  An interference pattern was  distinguishable in the
reflected beam profile, with a contrast and intensity varying with the
 position of the beam impact and with the incidence. We also observed that tuning the window temperature at a
reflection {\it maximum} ($R_{max} \simeq 20\%$) caused an increase of noise.  All the subsequent cells were
fabricated \cite{sar89} with windows made by Meller Optics \cite{meller} 
with both an excellent parallelism (better than 10$~\mu$rad over the 6 mm diameter central region) and very good [0001] crystal
axis orientation (defect on C-axis orientation
$\leq  0.5^{\circ}$, birefringence $\leq 2 \times 10^{-3}$ for a 0.5 mm thick window). As expected, we clearly
observed a reduction of the noise when the reflection was reduced to a few times $10^{-3} $ per window thanks to
temperature stabilization  (to 0.1$^{\circ}$ C), achieved independently for each window.  

Given the window thickness, we obtain successive reflection minima at 540 nm by shifting the
temperature by successive intervals of 14$^{\circ}$ C while the FSR corresponds to 40$^{\circ}$ C at the probe
wavelength. Hence, we cannot expect to obtain reflection extinction of comparable quality simultaneously at both
wavelengths. However, by varying the operating point by increments of
$14^{\circ}$ C  over a wide acceptable temperature range (210 -
270$^{\circ}$ C) one can choose an operating temperature which provides the  best probe transmission with
excellent reflection extinction for the excitation beam. In a majority of cells we could obtain a $\approx$
95$\%$ probe transmission leading to a further improvement of the observed SNR.
 This is likely to result from the quasi-suppression of the interference between the probe beams reflected at the input and output
windows. No systematic effect is expected from this interference, due to the way we process and
calibrate the PV data. As a result of drifts in the interference order, however, we do expect noise associated with  
temperature drifts of the cell-body, even though cell-windows are accurately stabilized. The smaller the amplitude
of this   interference, the smaller the associated noise.
 
 \subsubsection{ The tilt-odd effect: interpretation and suppression }
 When the probe beam is at normal incidence to one of the windows, we observe an excess of noise in the
polarimeter signals  making precise measurements impossible. Although the ideal configuration requires perfect
alignment of both laser beams along the cell axis, we have to concede a small tilt of the cell axis with respect to the
beams ($\psi
\sim 3$~mrad). Since such a tilt breaks the symmetry, we reverse its sign after about 90~minutes of data acquisition, and we
average the results obtained with both tilts, affected in practice by similar statistical noise. The reversal 
$\psi \leftrightarrow -\psi$ is performed by tilting the oven containing the cell while keeping the position of the beams 
unchanged. We now justify this procedure which suppresses a possible systematic effect. (An overview of
the systematics is given in Section IV).
 
 In reference \cite{bou03} (\S 4.4 Eq. 39), we showed that a misorientation of the probe beam with respect to the
excitation beam generates a second-order systematic effect on the measurement of $\theta^{pv}$ for a given
direction of $\hat \epsilon_{ex}$:
 \begin{eqnarray}
&\theta_{syst}(\hat \epsilon_{ex}) = \nonumber \\
&\hspace{-3mm}\frac {E_t^{+}}{E_l} \left( \hat k_{ex} \wedge \hat k_{pr}\cdot\hat E_t^{+} - (\hat k_{ex}
\wedge \hat k_{pr}\cdot
\hat \epsilon_{ex})(\hat E_t^{+} \cdot \hat \epsilon_{ex})\right) \hat z\cdot \hat E_l ~\; ~, 
 \end{eqnarray}     
 involving the alignment defect $\hat k_{ex} \wedge \hat k_{pr} = \delta
\alpha \, \hat n$ and the transverse $E_t^{+}$ field defect even under the longitudinal field reversal, both to first order.  
 In our experiment the pump-probe alignment
of the two beams is adjusted precisely enough, using a four-quadrant cell, to avoid any really significant
contribution. However, a problem can arise from the portion of the excitation beam {\it back-reflected} by the
output window, which is misaligned with the probe beam. 
Let us denote  by $\hat k'_{ex}$  the reflected beam direction, and $\hat n$ the direction such that $ \hat
k'_{ex}\wedge \hat k_{ex }= 2
\psi \, \hat n$, where $\psi $ is the angle of incidence.  We see that a contribution, linear in $\psi$, appears
in
$\theta_{syst}(\hat
\epsilon_{ex})$ (Eq. 3) that does not average to zero in the ``isotropic'' value, {\it  i.e.} the average after 90$^{\circ}$ rotation of
($\hat
\epsilon_{ex}, \hat \epsilon_{pr}$), \cite{bou03}:  
\begin{equation}
 <\theta_{syst}>_{\hat \epsilon_{ex}}=  \psi \,  R \, \frac {E_t^{+}}{E_l} \left( \hat n\cdot\hat E_t^{+} \right)\hat z\cdot \hat E_l \; ,
\end{equation}
where $R$ is the reflection coefficient of the output window.      
It simulates the PV effect, except that it is {\it odd} under the $\psi \leftrightarrow  -\psi$ reversal. We have
neglected the loss of detection efficiency due to the incomplete overlap of the probe beam and reflected excitation
beam since for
$\psi = 3$~mrad, the beam separation is only 0.5~mm at the entrance of the cell, which is less than the probe
beam radius of 0.7~mm.

With $R=0.10$, a typical value for standard sapphire (n=1.77) windows, taking $\psi $= 3~mrad and $E_t^{+}/E_l= 3 \times
10^{-3}$, using Eq. 4  we predict  $<\theta_{syst}>_{\hat \epsilon_{ex}} = 0.9 ~\mu$rad. This reduces to 0.02
$\mu$rad for good windows with $R$=0.002.
For larger tilts, the overlap of the probe and reflected excitation beams is partial and the  $\psi$-odd
contribution (Eq. 4) is not expected to grow linearly, but actually to saturate. Indeed, this corresponds to our
observations when the tilt is increased up to $\psi \sim $ 5mrad.  

In conclusion, the tilt-odd effect could correspond to a source of systematic effect if  there were no means to
suppress it.  Actually, we have
two ways to reduce this effect efficiently i) by reducing the reflection coefficient to the $10^{-3}$ level, and ii) by 
reversing the sign of $\psi $.
The most convenient way to perform the tilt reversal of the cell axis with respect to the unchanged common beam
direction, is to rotate the cell around a vertical axis passing through its center.  The displacement of the beam impact
on each window is only 240 $\mu$m. On such a small scale, the value of
$E_t^{+} $ is not expected to change, which is in fact confirmed by our control of the transverse fields. This is
important for efficient suppression of this effect.

We want to mention a second source of tilt-odd effect. In \cite{bou03} we analyzed the systematic effect
generated by  the combined action of the transverse electric and magnetic field components $E_t^{-}$ and
$B_t^{-}$, both  odd under reversal of the longitudinal field. By tilting the cell, together with the HV electrodes
assembly \cite{jah01}, we produce an electric component,
$E_t^{-}= \psi E_l$, and hence a $\psi$-odd systematic effect. Using Eq. 34 of ref. \cite{bou03}, we obtain the new
isotropic contribution to $<\theta_{syst}>_{ex}  $:
$$<\theta_{syst}>_{\epsilon_{ex}} = (\hat z \cdot \hat E_l)  \psi \,  \omega_{_{F'}} \tau \, (\hat E_t^{-} 
\cdot \hat B_t^{-})\;.    $$

For typical values of $B_t^{-}$ less than 2~mG (leading to a Larmor precession angle of 
$ \omega_{_{F'}} \tau\leq 40 ~\mu$rad) and $\psi
= 3$~mrad, we obtain $ <\theta_{syst}>_{\epsilon_{ex}} \leq  0.12 ~ \mu{\rm rad}$.  
 Thanks to the suppression of $ <\theta_{syst}>_{\epsilon_{ex}}$  in the  $\psi \leftrightarrow -\psi$ reversal, we
can consider this effect to be  harmless.     
  
\subsection{A cesium cell with electrical continuity between inner and outer sides of conductive windows}
 The last improvement consisted in our using sapphire windows covered with a niobium, 2~$\mu$m thick, coating
deposited over a thin layer of titanium, except for the 6~mm central region left uncovered for the laser beams. This
type of coating allows one to control better the electric field near the windows, since the HV potential can be applied
{\it inside} the cesium cell by direct contact with the {\it outer}  part of the metal coating. By comparison with what
happens when the window potential is floating, the electric charges left at the windows as a result of photoionization
are more efficiently compensated by those supplied by the generator maintaining the potential fixed.

The vacuum-tight gluing of such coated windows to the alumina tube, under vacuum to prevent the coating 
from oxidation, was implemented by David Sarkisyan and co-workers \cite{sar04}.  
  
 In this last cell we have obtained the best SNR, corresponding to a further $\sim 15 \%$ improvement.

A measurement method relying on atomic signals (described
in Sec. V D) has allowed us to determine precisely the electric field
experienced by the atoms inside the cell. It is interesting to compare the measured value 
$E_l^{exp}$ with the magnitude $E^{nom}$ expected from the numerical simulation taking into account the
geometry of the electrode assembly and the potential distribution (for
details see \cite{jah01}).  The results exhibit a marked difference between the cell having electrically conductive
windows,  where we find $E_l^{exp}/E^{nom} = 0.98 \pm 0.01$, and a cell with uncoated windows, leading to
$E_l^{exp}/E^{nom} = 0.92
\pm 0.01$. The simulation does not take into account the distribution of electric charges inside the cell
resulting from the photoionization at the windows. Accounting for this process, the observed variation of 
$E_l^{exp}/E^{nom}$ from one type of cell to the other is not surprising: photo-emission leaves a positive surface charge at
the cathode window, and the accelerated electrons accumulate at the anode window, giving rise to a negative surface charge.
Applying the potential, via the coatings, at the inside surface of the windows contributes to screening the effect of the surface
charges in the vapor. In addition, the photoelectrons that reach the anode window at the periphery are evacuated through the
coating. The applied electric field is then expected to be closer to the calculated electrode field.

\subsection{Net observed improvement of the S/N ratio} 

After this set of improvements (\S A to E), compared to our initial runs 
reported in \cite{bou02}, the average value of the standard deviation per isotropic value of the calibrated $E_l$-odd linear
dichroism (see Sect. V.A) has been
reduced by a factor of 2.6 (initially 5.1~$\mu$rad and now
2.0~$\mu$rad) while the repetition rate has been increased from 90 to 160~Hz. All in all, the SNR for a 
one-second-measurement time has been increased by a
factor of $\sim 3.5$, hence {\it the averaging time required to reach a given statistical accuracy is
reduced by a factor of  $\sim$12}, and this without introducing new systematic or spurious effects.

  Even so, this does not correspond to a technological limit: for
instance, the same kind of pulsed laser we are using has been operated at a repetiton rate reaching 400 Hz
\cite{liz}. This would provide  another improvement by a factor of  $\sim$~1.5, provided that the pointing
stability of the excitation beam can be preserved when the repetition rate is doubled.

We have already discussed the noise-equivalent-angle and shown \cite{cha97,bou05} that, for a given number of
incident probe photons,
$n_{in}$, the quantum noise limited SNR per excitation pulse,
$$SNR =\theta^{pv} \sqrt n_{in}\, {\cal A}_{av} \exp{({\cal A}_{av}/2)} \, ,   $$ is a rapidly growing function of
the optical density averaged over the para and ortho configurations, ${\cal A}_{av} = ({\cal A}_{\parallel} + {\cal
A}_{\perp})/2$. Actually our measurements show an excess of noise with respect to the shot noise limit by a factor
of 1.5 to 2. Even so, while making the various  improvements of our experiment, we have checked that the
mechanism of asymmetry amplification by stimulated emission is in practice a definite source of improvement of the
SNR. The optical density involves the number of excited atoms in the vapor column through which passes the probe
beam,
$N_{ex}$. The latter is proportional to $E_l^2$ and to the excitation intensity. We have increased those parameters,
though without overstepping the limits beyond which new sources of noise might arise. This is especially
important when we increase the electric field (see Sect. V. B our diagnosis of noise at short time scale).  In practice,
for the $6S_{F=3}
\rightarrow 7S_{F=4} \rightarrow 6P_{3/2,F=4} $ transition, good measurement conditions were achieved with
typical values of the asymmetry $A_{LR}/\theta$  of 1.2 (resp. 0.6), at optical densities ${\cal{A}}$ of  0.5 (resp. 1.0) in the ortho
(resp. para) configurations.  

%%%%%%%%%%%
\section{Control, reduction and estimation of the systematic effects}
%%%%%%%%%%
In this section, we present an overview of the origins of the systematic effects and the means we adopted to reduce and
estimate them,  (to the exception of the tilt-odd effects considered in Sect. III D) . The order chosen in this
presentation corresponds to decreasing order of importance played by each effect.
   
 \subsection{ The longitudinal magnetic field odd under reversal of  $\vec E_l$}

We have observed a longitudinal magnetic field odd under reversal of  $\vec E_l$, dubbed the $B_z^{-}$ field. Its 
likely origin is the motion of electric charges following the photoionization process which may have a small
helicity around the propagation axis. The $B_z^{-}$ field gives rise to an  $E_l $-odd Larmor precession of the axes of
the parity-conserving linear dichroism, thus simulating the PV tilt angle.  Even in a field as small as $B_z^{-} = 50
~ \mu$G, the precession is of the same order of magnitude as $\theta^{pv}$.    
 
We measure this field by observing the optical rotation, 
{\it odd} under $\vec E_l$ reversal, that it generates by a simple Faraday effect \cite{bou95}. This control is
performed  by selecting a hyperfine component  of the probe transition particularly sensitive to a magnetic field.
The Faraday effect  on the
$7S_{F=4}
\rightarrow 6P_{3/2,F=5}$ line is 10 times larger than the linear dichroism resulting from the Larmor precession of the excited
state alignment which might simulate APV. In order not to rely on the temporal stability of the value of  $B_z^{-}$ during 
long acquisition times, the measurement is made before and after the PV data taking. At regular intervals the  calibration
factor for the Faraday effect is obtained by applying a ``large'', known, magnetic field. This is also done on the
$7S_{F=4}
\rightarrow 6P_{3/2,F=4}$ probe transition for the calibration factor of the dichroism precession. Thus the measured value of
$\theta^{pv}$  can be corrected for reliably and accurately. Both the sign and magnitude of  $B_z^{-}$ field varied from one
cell to another. In the best cases, the correction remained at the level of a few percent of the PV effect, while it was of the same
order of magnitude in two cells (cells $\#$5 and 6). The time devoted to
$B_z^{-}$ measurements varied from 30 to 60$\%$ of the total data acquisition time. It was chosen so that the
error associated with the resulting correction on $\theta^{pv}$ remained small compared to the statistical accuracy of the PV
measurement.

\subsection{Effects resulting from a breaking of the cylindrical symmetry}

 Particular attention has been given to the defects that break
the cylindrical symmetry of the set-up, such as transverse $\vec E$ and $\vec B $ fields and misalignment of the two
beams. Our study
\cite{bou03} has shown that for the polarimeter imbalance to be altered in a way which simulates $\theta^{pv}$,
 {\it two defects} are necessary. Some of the systematics (``class 2''-systematics in \cite{bou03})
average to zero when the two polarizations $\hat \epsilon_{ex}, \hat
\epsilon_{pr}$ are rotated together by
$45^{\circ}$ increments around the common beam direction. The really serious effects (``class 1'') are those that do
not average to zero under this operation. They all require the presence of a transverse electric field. 

These ``class 1'' effects have two different origins:
  
 \hspace{1.5mm}

\noindent - 1. Pump-probe misalignment of angle $\delta \alpha \, \hat n = \hat k_{ex} \wedge \hat k_{pr}$.

\noindent  This misalignment gives rise to a systematic effect on the measurement of $\theta^{pv}$ by its coupling to
an $E_t^{+}$ electric field, even under $E_l$ reversal. Our method to minimize this effect is: 

                \hspace{5mm} i) to superpose the pump and probe beams, on the {\it same centering device}, at the input
and output of the cell, (see Sect.  3.3 in ref \cite{gue98});
 
                \hspace{5mm} ii) to measure the transverse $E_t^{+}$ field (procedure in next
section), and then slightly translate the cell transversally  along $x$ or $y$ so as to reduce it to the level of
$\sim$ 1~V/cm, knowing from
previous studies \cite{gue02} that $E_t^{+}$ has a centripetal distribution around the cell axis. 

 \hspace{1.5mm}

\noindent - 2. Coupling of a transverse $\vec E_t$ and a transverse $\vec B_t$ magnetic field

\noindent A ``class 1'' systematic effect can also arise from $\vec E_t^{+} \,, \vec B_t^{+}$ or $\vec E_t^{-} \, , \vec B_t^{-}$
couplings. For this reason it is necessary to measure (and, as much as   possible, to minimize) the values of $ B_x^{+}\, , B_x^{-} \, ,
 B_y^{+}$ and $ B_y^{-}$, as well as
$ E_x^{+}, E_x^{-}, E_y^{+}$ and $ E_y^{-}$. This is achieved, with the probe tuned to the $7S_{1/2,F=4} \rightarrow
6P_{3/2,F=5}$ hyperfine component, by performing sequences of measurements similar to the PV sequences, except
that a ``large'' transverse magnetic field (1~G) is applied and reversed, along $x$ then along $y$. The second-order magnetic
perturbation of the Stark dichroism of well-defined signature allows us to extract the components of the transverse 
magnetic fields (see Sect.  5 in
\cite{bou03}), while we exploit an optical rotation signal to extract the transverse electric fields \cite{bou05b}.

\hspace{1.5mm}

Table II summarizes the means we use to minimize the field defects. From day to day, only the $ \vec B_t^{+}$
field needs to be readjusted in order to be kept at the 2~mG level.  The duration of this control is negligible 
compared with the  data acquisition time needed for measuring $B_z^{-}$. The same control procedure is
performed at the beginning and at the end of the PV sequences on the $7S_{1/2,F=4} \rightarrow
6P_{3/2,F=4}$ transition, for both tilts $\pm \psi$ of the cell. These measurements
allow us to evaluate the systematics per milligauss of stray
$\vec B_t$-field components. The measured values are then combined with the residual $\vec B_t$ field values  
extracted on the  $7S_{1/2,F=4} \rightarrow
6P_{3/2,F=5}$ transition (for the 
same tilt of the cell), to yield the systematics affecting the PV data. On a day-to-day basis, the size of these effects 
is a few percent of the PV
effect. They are affected by a statistical uncertainty small compared to the statistical error on $\theta^{pv}$ and could be
corrected for
 when significantly non-zero. But, on the average for a given cell, these effects was kept below the percent 
level, with the exception of one cell (cell $\# 4$) for which a correction of $\approx$ 10$\%$ was applied to one
third of the data with its uncertainty taken into account. We have no indication that transverse field effects might be a
major problem for a future 1$\%$ precision  measurement. 

 Besides the estimation of the effects that break the cylindrical symmetry, a test of isotropy on the
PV data themselves provides a diagnosis of their presence (see Sect. V C). 
  
\begin{table}
\caption{Means of reduction of the $E_l$-odd  and the   $E_l$-{even} components of the transverse $\vec E_t$ and $\vec
B_t$ fields.     }
\vspace{5mm}
\begin{tabular}{|c |c|c |}
\hline Defect & Origin & Reduction \\ \hline

$E_t^{-}$& Tilt of the cell & ($\psi/-\psi$) mean tilt \\ \hline
$E_t^{+}, B_t^{-}$& Photoemitted charges \cite{gue02} &  Cell translation $\perp \vec E_l$ \\ \hline
$B_t^{+}$& Residual ambient field & Compensating coils \\ \hline

\end{tabular}\\
\end{table}
 
\subsection{Possible instrumental defect affecting the orientation of $\hat \epsilon_{ex}$}
 We have considered the possible existence of a tilt $\theta_0 ^{-}$, odd under $\vec E_l$ reversal, affecting the
excitation polarization at the entrance of the Cs cell.
 Since the direction of $\hat \epsilon_{ex}$ determines the direction of the P-conserving gain axes, such an instrumental defect would
exactly simulate the PV tilt $\theta^{pv}$. It is therefore crucial to check that the direction of 
$\hat \epsilon_{ex}$ is unaffected by the field reversal. During PV data acquisition, a second polarimeter is used
to analyze the excitation polarization using a fraction of the main beam, picked off at the cell entrance (see Fig 3).
Throughout the measurements, $\theta_0^{-}$ remained at or below the noise level, and the global result, 
$\theta_0^{-} $ = - 0.030 $\pm ~0.020~\mu$rad, is compatible with zero. This kind of effect might have arisen from
electromagnetic interferences resulting from pulsed $\vec E_l$ operation.  

\subsection{Misreversal of $\vec E_l$ combined with polarization defects}
A misreversal of $\vec E_l$ cannot contribute by itself, but only through a combined effect also involving polarization
defects ({\it e.g.} imperfect  parallelism of $\hat \epsilon_{ex}$ and $\hat \epsilon_{pr}$). In fact our reconstitution
method protects us efficiently against such an effect  since we perform the imbalance calibration for both signs of the
$\vec E_l $ field. As one can check from Eq. (8) below, this method eliminates any field misreversal from the outset.
Nevertheless, the defects are kept below the noise level: field misreversal $\leq 10^{-3}$ with a digital servo loop,
and polarization imperfections $\leq 10^{-4}$ by preliminary manual corrections based on atomic signals
\cite{gue97} and real-time monitoring. The defects are stable owing to the good optical quality, homogeneity, small
birefringence..., of the optical components and cell windows, as well as good reproducibility  of the insertion of the
$\lambda/2$ plates.    

%%%%%%%%%%%%%%%%%%
\section{Data acquisition and Processing. Calibrations. Results}
%%%%%%%%%%%%%%%%%%
\subsection{Reconstruction of the PV signal and data acquisition sequences}
On an excitation pulse basis, our dual channel polarimeter provides the imbalance, $(S_1-S_2)/(S_1 + S_2)$, and the probe
intensity $I = S_1 + S_2$ for both the amplified and reference pulses respectively.
From these signals, two main quantities are formed: the asymmetry
\begin{equation}
 A_{LR} \equiv D_{at} = D_{amp} - D_{ref}\, ,
\end{equation}  
and the optical density for the probe,
\begin{equation}
  ln{ (I_{amp}/I_{ref}) } = {\cal{A}} + {\cal{A}}_0 \; .
\end{equation} 
In Eq. 6, ${\cal{A}}_0$ is a small negative contribution to the optical density  due to {\it absorption} of the probe beam by a $6P$
population of known collisional origin 
\cite{note6P}. It is measured once at the beginning of data taking, with the excitation beam detuned a few gigahertz
away from the forbidden transition.  ${\cal{A}}_0$ typically amounts to $-4\times 10^{-2}$, whereas ${\cal{A}} \simeq 1$ at
resonance in the para-polarization configuration. No such background is detected on $A_{LR} $. 
 
 We also form the ratio  
\begin{equation}
\vartheta = A_{LR}/2 \lbrack\exp{(\eta {\cal{A})}-1)}\rbrack \,.   
\end{equation}
  This ratio just provides the tilt angle
$\theta$  of the eigenaxes (Eq. 1) within a normalization factor close to 1 (or close to the magnification factor when the
polarization magnifier is used), eliminated in the calibration procedure measuring the same quantity for the known
$\theta_{cal}$ angle. Then over the 30 consecutive laser shots corresponding to a given state of the experiment,
the program estimates both the means and standard deviations of $\vartheta$ and $A_{LR} $ and stores them for
 the purpose of later analysis. This sequence is repeated for all the states depicted in Table 1. For each
reversal, the sign of the initial state $\sigma_{cal}, \sigma_{_E}, \sigma_{det},
\sigma_{pr} = \pm1$ is chosen at random. The complete signature of the APV
signal involves an average over all the $2^4$ possible states. For a given $f(\sigma$), we define the average, $< f(\sigma)>_{\sigma} =
\frac{1}{2}(f(1) + f(-1))$, {\it i.e.} the $\sigma$-{even} part. This implies that  $<  \sigma f(\sigma)>_{\sigma} =
\frac{1}{2}(f(1) - f(-1))$ yields the $\sigma$-{odd} part. The first determination of the PV  calibrated linear
dichroism in a given {\it excitation} polarization state, $i = \lbrace x, y, u, v \rbrace$, involves the construction of the
following quantity:
\begin{eqnarray}
< G_i >_{A_{LR}} =  \hspace{60mm}\nonumber \\
\theta_{cal} \left <   \sigma_{_E} \left  [\frac{ <
\sigma_{det} A_{LR} (\lbrace \sigma_j  \rbrace) >_{\sigma_{det} \sigma_{cal}}}{< \sigma_{det}
\sigma_{cal}A_{LR}(\lbrace \sigma_j \rbrace ) > _{\sigma_{det}
\sigma_{cal}}} \right ]   \right >_{ \sigma_{_E} \sigma_{pr}} .   
\end{eqnarray}
\begingroup
\squeezetable 
\begin{table}
\caption{
 Main sequences of data acquisition involved in a run, listed in chronological order }
\begin{center}
\begin{tabular}{|p{1.0cm}|p{2.1cm} | p{3.7cm}|p{1.3cm}|}

\hline ${\rm Tilt~of}$ ${\rm the~cell}$ & ${\rm Probe~hfs~line}$ ~~~~$~~7S_{F} - 6P_{3/2,F^{'}}$ &  Type of measurement & 
Duration
\\ 
\hline
$~~~\psi$ & $F=4-F^{'}=5$ & $B_z^{-}$measurement{\rm ~PV-type} & $\sim$ 60 min\\  
&~& ${\rm \&~transverse ~field~control}$&~~~~5 min\\ \hline
$~~~ \psi$ & $F=4-F^{'}=4$ &  PV$-$measurement & $\sim$ 90 min \\ 
&~& ${\rm \&~transverse ~field~control}$&~~~~5 min\\\hline
$ - \psi$ & $F=4-F^{'}=4$ &  PV$-$measurement & $\sim$ 90 min \\
&~& ${\rm \&~transverse ~field~control}$&~~~~5 min\\ \hline
$ - \psi$ & $F=4-F^{'}=5$ & $B_z^{-}$measurement{\rm~PV-type} & $\sim$ 60 min \\
&~& ${\rm \&~transverse ~field~control}$&~~~~5 min\\ \hline

\end{tabular}
\end{center}
\end{table}
\endgroup  

A second determination $< G_i >_{\vartheta}$ is obtained by replacing in the above expression $A_{LR}$ by $\vartheta$ (Eq. 7).
Both methods give identical results.   
Note that an imperfect $\vec E_l$ reversal alone leading to an $\vec E_{l}$-odd
contribution to ${\cal A}$ does not affect either quantity, thus ensuring suppression of systematics due to $\vec
E_l$-misreversal. However, it can be shown that a contribution to ${\cal{A}}$, that is both $E_l$-odd and
$\theta_{cal}$-odd  does not affect $< G_i >_{A_{LR}}$ whereas it would alter
$< G_i >_{\theta}$. In this particular example, we see that it is instructive to compare the results obtained with
both methods.   From the point of view of SNR, we noted that in the method using $\vartheta$, it is
somewhat more advantageous to evaluate $\vartheta$ over one state (30 laser shots) by replacing ${\cal{A}}$ in Eq. (7) 
by its average over the 30 laser shots:  
$\overline \vartheta = \left <  A_{LR}  / 2 (\exp {\eta \left < {\cal{A}} \right > -1 ) } \right > $.
 
Each four-polarization cycle yields two ``isotropic values'':  $S_{xy} = \frac {1}{2} (G_x + G_y) $
and $S_{uv} = \frac {1}{2} (G_u + G_v) $. The polarization cycles are repeated over $\sim$ 90~min, providing us
with
$N_{iso}$ ``isotropic values'' (typically
$N_{iso}$=30) for a given tilt $\psi$ of the cell and repeated for the opposite tilt $ -\psi$. For each tilt, the
$B_z^{-}$ correction is also determined as well as the sensitivity to $B_t$-dependent systematic effects. The main
data acquisition sequences are summarized in Table~III. This constitutes a so-called run
$\# k$, providing us with an ensemble of $N_k = 2 N_{iso}$ PV data. From this ensemble, we deduce an average
value $m_k$ and the standard error $\sigma_k$. The average $m_k$ represents the calibrated tilt angle of the gain
axes having the complete PV signature defined in Table~I, {\it i.e.} $\theta^{pv}_{exp}$. We accumulated typically 30 such runs
using a given cell. The run results are merged with weights $1/\sigma_k^2$ to give a single result per cell. Alternatively, the
$N_k$ individual PV data of all runs of all cells are merged into a single ensemble, the same weight being attributed to
any individual datum.
 The results are presented and compared in Sect. V D. 

Besides the PV quantities, several other quantities bearing non-PV signatures are constructed 
from the polarimeter signals, providing us on a short time-scale with a wealth of information making possible
real-time corrections for defects or drifts during lengthy data acquisition. The most important of these are:
 
 $-$  ${\cal{A}} \; {\rm and} \;A_{LR}(\theta_{cal}$-odd), which should be kept maximum since they condition the sensitivity (a
decrease is generally due to a drop of $I_{ex}$).

 $-$ Asymmetries under $\vec E_l $ reversal of  ${\cal{A}}$ and $A_{LR}(\theta_{cal}$-odd), exploited to cancel the field
misreversal.

 $-$ Reference imbalance of the polarimeter revealing probe polarization defects and/or drift of the difference of the gains between
the two channels.

 $-$ Atomic imbalances: i) $E_l$-even and $\sigma_{det}$-odd reveal pump-probe polarization defects; ii)  $E_l$-odd and 
$\sigma_{det}$-even reveal parasitic  electrical noise.

 \subsection{Noise peak rejection and test of PV data rejection} 
The aim of noise peak rejection is to discard accidental outlying data without truncating the noise distribution. 
During data acquisition noise blasts which can affect $D_{at}$ are immediately detected:
if the standard deviation  of the atomic imbalance  $\sigma_{D_{at}}$, estimated in one state of the experiment ({\it i.e.} over 30
laser shots), happens to exceed three times its typical  value,  the corresponding measurement is ignored and immediately
repeated {\it before the next parameter reversal is performed}. Under normal conditions, such brief events occur with a
probability of only a fraction of a percent. In a few cases, this precaution proved useful to eliminate noise not continuously present,
but possibly associated with the electric field shots. This was the case of cell $\#$ 3 mentioned later on, in Sect. V.
E.)
 
In  time-deferred analysis, we eliminate outlying PV data by self consistent truncation at three standard deviations on the
distribution of the PV data accumulated in a given tilt of the cell. The number of rejected data is small ($\leq
1\%$, barely larger than what is expected for a standard gaussian distribution, 2.6$\times 10^{-3}$). 
Avoided laser-mode hops, 
or imperfect plate positioning after insertion are typical possible causes for outlying data. Since the noise
distribution is expected and observed to be symmetric this truncation operation introduces no bias  on the data and
reduces slightly the standard deviation. It was also performed over the
distributions of PV data at the various stages of the analysis, {\it i.e.} over the runs in a given cell and over all data
merged together. 
 
\subsection{Test of Isotropy}
\begin{figure*}
 \centerline{\epsfxsize=150mm  \epsfbox{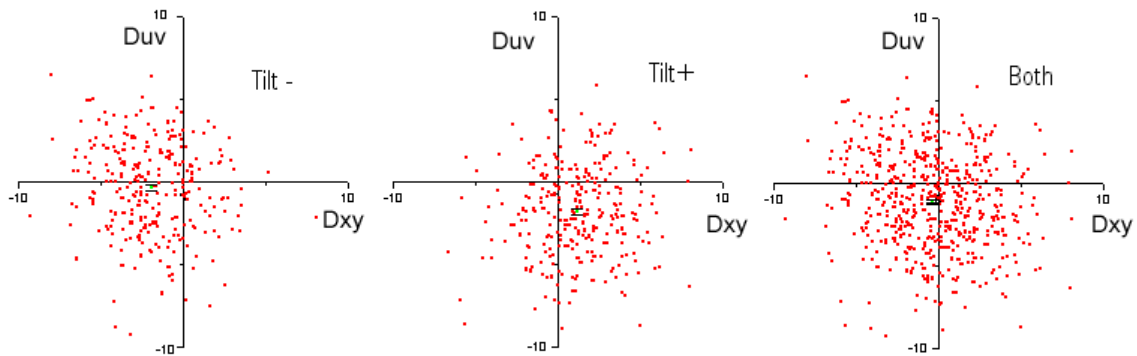}}
 \caption{ Results of the anisotropy test performed on an ensemble of 1126 data points obtained in
cell \#4. The signals $D_{xy}$ and $D_{uv}$ are analyzed either separately according to the sign the tilt $\psi$ of
the cell (two graphs on the left), or altogether (graph on the right). In this data set there is no tilt-odd contribution to
$S_{xy}$ nor $S_{uv}$ coming out of noise. See text for the interpretation.}
 \end{figure*}
 The two values of the isotropic part of the $E_l$-odd linear dichroism, $S_{xy} $
and $S_{uv} $, that we extract in each four-polarization cycle   
  are found compatible within the noise level, 
as one expects from considerations of the symmetry of the Stark dipole of the
excitation transition \cite{bou03}. This is observed on individual runs but it is also confirmed by the global analysis
of all data (see Eq. 11 below). The presence of defects breaking the cylindrical symmetry, responsible for {\it both}
``class1'' {\it and} ``class 2'' systematics is expected to show up as non-zero differences $
D_{xy}=\frac{1}{2}(G_x-G_y)$ and $D_{uv} =
\frac{1}{2}(G_u-G_v)$. The isotropy test consists in plotting one point of coordinates $(D_{xy}, D_{uv})$ per data
set in a cartesian coordinate system. In conditions of perfect isotropy, the center of gravity of the cloud of points
should merge into the origin within the error bars. In
\cite{bou03} we presented a set of data presenting no significant anisotropy. Here, we present (Fig. 10) another
set (1126 data points obtained in cell \#4) analyzed separately for the two opposite signs of the tilt. The anisotropy
is clearly apparent in each tilt, with $D_{xy}$ signals of opposite signs. It is reduced over the whole data set. By
contrast, the isotropic contributions in both tilts are statistically compatible.  In other words $D_{xy}$ appears much
more sensitive than $S_{xy}$ or $S_{uv}$  to the anisotropy induced by the tilt. Referring to \cite{bou03}, we expect
the tilt of the cell to give rise to effects of both ``class~1'' and ``class~2'', but   
 for a tilt $\psi_x$ (\i.e. a rotation around $\hat y$), ``class~1'' effects would contribute to both $D_{xy}$ and the
isotropic parts, $S_{xy}$ and $S_{uv}$, with comparable magnitudes (Eqs. 35, 37, 39 40 in \cite{bou03}). On
the other hand,  tilt-dependent ``class~2'' effects cancel out in $S_{xy}$ and $S_{uv}$, hence are not a source of systematic 
effect, but can contribute to both $D_{xy}$  and
$D_{uv}$.  On this particular data set, we interpret the value of  $D_{uv}$ remaining after averaging over the two
tilts (Fig. 10, right graph) by the presence of a residual tilt $\psi_y$, independent of the
$\psi_x / -\psi_x$ reversal. Therefore, this signal is a useful warning indicating a defect but is not the sign of a systematic effect.   

Moreover, when the isotropy test is performed on the whole data set, from one cell to another residual anisotropies
tend to compensate. This means that their principal origin is not in the optical components of the set-up but rather
arises from slight residual imperfections occuring either during the mounting of each cell inside the electrode
assembly or during the fabrication of each individual cell. However, as shown by the final results and the discussion
presented hereafter,  there is at present no hint of any  significant residual systematic effect varying from one cell to
the next.           
%%%%%%%%%%
\subsection{Measurement  of $\vec E_l$ and $\theta_{cal}$ for calibrating $\theta^{pv}$}
%%%%%%%%%%%%
\subsubsection{A precise {\it in situ} measurement of $\vec E_l$}
To take advantage of the substantial reduction achieved in the statistical uncertainty (see Sect. IV F), we were obliged to reduce 
also the uncertainty on the magnitude $E_l$ of the field inside the cell, this value being required for a
comparison of 
 experiment with theory. For this purpose we changed our calibration method \cite{bou02}.
It can now be conducted in  the exact conditions of PV data acquisition: same hyperfine probe transition, excitation energy and
applied potentials.  It provides us with reliable results to within a one percent accuracy.

The basic idea relies on the comparison of two optical densities of the vapor at the probe wavelength, the first without any
applied  electric field and the second in the longitudinal field of magnitude $E_l$ to be measured. They are both   
  proportional to the number of atoms  excited in the $7S$ state, hence to the excitation probability,
respectively  
$M^{'2}_1 $ and $\beta^2 E_l^2 + M^{'2}_1$ to within identical proportionality factors, ($M^{'}_1$ denotes 
the $6S-7S$ transition amplitude\cite{noteM}). From the optical density ratio we can thus deduce $M^{'2}_1/\beta^2
E_l^2$, {\it i.e.}
$E_l
$ in terms of the precisely known atomic  quantity 
$M^{'}_1/\beta = 35.1 \pm 0.1 $~V/cm \cite{bou97,bou88,ben99}.  The optical density is deduced from the polarimeter imbalance
resulting from the  left-right asymmetry
 $A_{LR}$  that is associated with the $7S$ atomic alignment arising from a tilt
$\theta$ of $\hat \epsilon_{ex}$ with respect to
$\hat \epsilon_{pr}$, ({\it i.e.} similar to the calibration signal used for data acquisition). The relation connecting
the optical density to $A_{LR}$ can be established precisely by relying on theory \cite{bou96}.
  
  In a first approximation, the result is given by the simple analytical expression,
$A_{LR} =  2 \theta \;\lbrack\exp{(\eta_{\perp} {\cal{A}_{\perp})}-1}\rbrack$ 
supposing  $\theta \ll
1$, and the ortho configuration with $\eta_{\perp} = 11/12$ for the
$6S_{F=3}-7S_{F=4}-6P_{3/2,F=4}$ transition. However, this result  is rigorously valid only if one assumes a probe
pulse duration $t_p$ long compared with the decay time  $\gamma_d^{-1}$ of the $7S-6P_{3/2}$  
optical dipole and short  compared with the    
$7S$ lifetime.  
Actually, in the real conditions of our experiment ($\gamma_d^{-1} = 13.4$~ns, $t_p = 20$~ns, $\tau_{7S} = 47.5$~ns) the
deviation with respect to the exponential amplification model although relatively small ($\leq$ 10$\%$), is non-negligible in
view of the precision sought. The exact result is deduced from a numerical solution of the exact equations derived in
\cite{bou96} (see Appendix B), using  efficient subroutines provided by Mathematica
\cite{wolf}. 

We have an important reason for choosing the linear dichroism resulting from the $7S$ atomic alignment as the observable
quantity to obtain the value of  ${\cal{A}}$ rather than the more direct determination (cf. Eq. 6) obtained from 
$ln(I_{amp}/I_{ref})$:
  in a zero electric field, differential
measurements providing the asymmetry
$A_{LR}$ can be performed accurately, while in the same conditions $ {\cal{A}}$ (of the order of $ 2 \times
10^{-4}$) is overwhelmed by noise. Moreover we observe no background superimposed on the atomic 
alignment ($< 0.3\%$ of the alignment in zero electric field). For this reason, observing the atomic alignment instead of 
the orientation created by a circularly polarized excitation beam as we did previously \cite{bou02}, corresponds to
a real improvement.  In addition, for the linear dichroism signal detected on the
$6S_{F=3}-7S_{F=4}-6P_{3/2,F=4}$ transition, our  experimental results confirm that saturation effects are
 especially weak, as expected in \cite{cha98}.          

For practical reasons, for the zero-field measurements (weak optical density), we adjust $\hat \epsilon_{ex}$
at 45$^\circ$ from
$\hat \epsilon_{pr}$ to detect the maximum value of the asymmetry, $A_{LR} = \eta_{\perp}
{\cal{A}}_{\perp}$, while in the
$E_l$ field measurement
$\hat \epsilon_{ex}$ deviates from
$\vec E_l \wedge \hat \epsilon_{pr}$ by a small known angle, $\theta = 12.41 \pm 0.06$~mrad, sequentially reversed from $+
\theta
$ to $-\theta$. In this way, the imbalance ratio to be measured is of order 70, even though the optial 
densities differ by a factor of $\sim$ 2000. In this way, we completely avoid possible non-linearity problems in the
detection chain. To eliminate saturation effects, measurements are performed at different levels of the probe beam intensity to
allow for an extrapolation to zero intensity, both with and without the applied electric field since saturation effects
depend on the amplification level. The field magnitude is then determined using: 
\begin{equation}
E_l^{exp}=  \frac{M^{'}_1}{\beta} \left (  \sqrt{ \frac{\ln{\lbrack 1+  A_{LR}(E=E_l )/2\theta} \rbrack }{(1+ \epsilon)
A_{LR}(E=0,
\theta= \pi/4)}}  -1 \right )  \,. 
\end{equation}
Here $\epsilon$ is the small quantity expressing the deviation of the exact result with respect to the simple one
assuming an exponential-type amplification; it is a function of ${\cal{A}}_{\perp}(E_l)$
depending on the value taken by the parameter
$\gamma_d t_p$. For example for the realistic values $\gamma_d t_p = 1.49$, and ${\cal{A}}_{\perp}(E_l) = 0.68$, we obtain 
$\epsilon=0.100 $.
 The precision in $\epsilon$ is limited by the uncertainty on $\gamma_d$, itself a linear function of the cesium atomic
density \cite{lin89}. Allowing for $5\%$ uncertainty on this latter, hence 3$\%$ one on $\gamma_d$,  the
resulting uncertainty on $E_l^{exp}$ is $0.3\%$.  In practical conditions, the precision in $E_l$ is that of the measurements,
presently better than 1$\%$. The determination of
$E_l$ by this method has been performed in the two types of Cs cells with this level of precision, leading to the results discussed in
Sect. III. E.
 The uncertainty on $E_l$ is negligible in comparison with the statistical uncertainty on
$\theta^{pv}_{exp}$ achieved in each type of cell. 
In a future work, we plan to investigate the limitations to the signal interpretation which may arise if one wants to push
further the precision of this method. 
  
\subsubsection{Calibration of the polarization-tilt angles}
Our measurements of $\theta^{pv}$ as well as those of the electric field suppose a precise knowledge of the tilt angles
realized by the Faraday rotator in terms of the applied current. Therefore, the calibration of the modulation
angle versus the applied Faraday current was repeated several times during the course of our PV measurements. It
is done by measuring the mechanical rotation of a Glan prism assembled on a precisely graduated
mount  which compensates the Faraday rotation. The precision of this calibration, 
0.5~$\%$, could be improved if need be.

%%%%%%%%%%%%
\subsection{Results}
\begin{figure}
\centerline{\epsfxsize=80mm \epsfbox{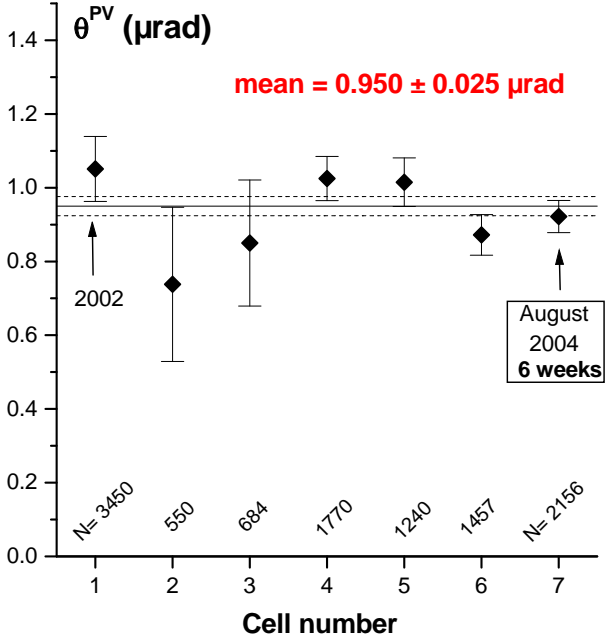}}  
\caption{\label{rescel}  Experimental values of $\theta ^{pv}_{exp}$($\mu$rad) obtained in different
cells, with their statistical error and the number of isotropic values accumulated in each cell to obtain the result. The solid
(respectively, dashed) line represents the mean (respectively, the statistical error on this mean). }
\end{figure}

\begin{figure}
\centerline{\epsfxsize=70mm \epsfbox{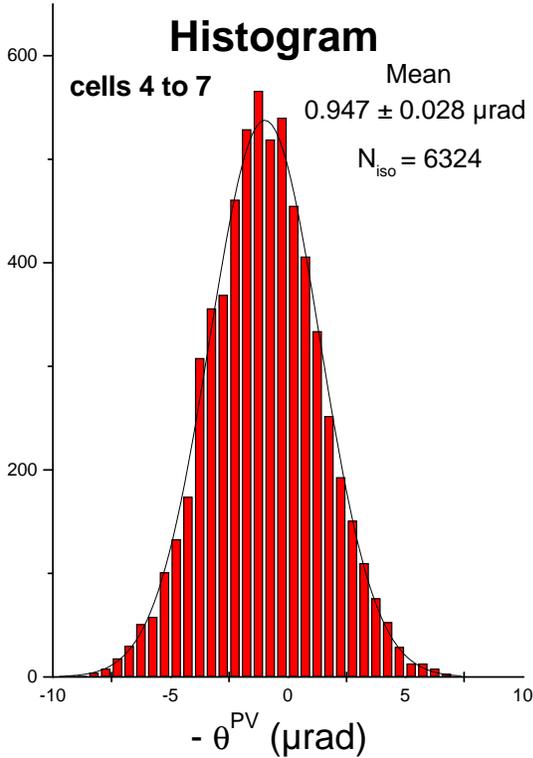}}
\caption{\label{histo} Right: Histogram of all the data accumulated in
cells 4 to 7 which have a largely dominant statistical weight. The line represents the gaussian distribution which has the same
mean and standard deviation. }
\end{figure}

\begin{figure}
\centerline{\epsfxsize=88mm \epsfbox{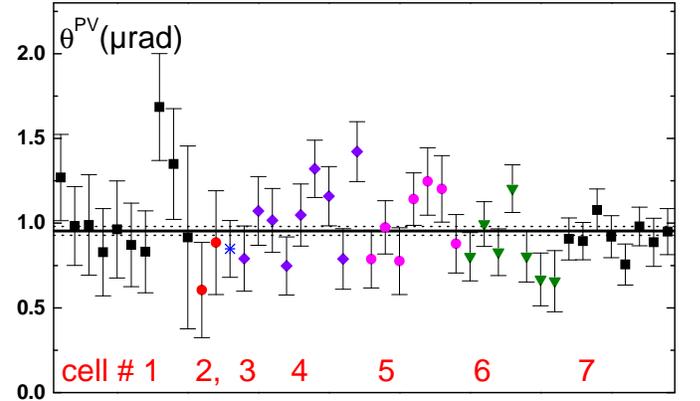}} 
\caption{\label{detres} Experimental values of  $\theta ^{pv}_{exp}$($\mu$rad) with their statistical error obtained in
successively accumulated runs (each point corresponds to $\simeq$ 4 runs), plotted versus the cell numbers chronologically
ordered. The solid (respectively, dashed) line represents the global mean (respectively, the statistical error on this mean).}  
\end{figure}

\begin{figure}
\centerline{\epsfxsize=70mm \epsfbox{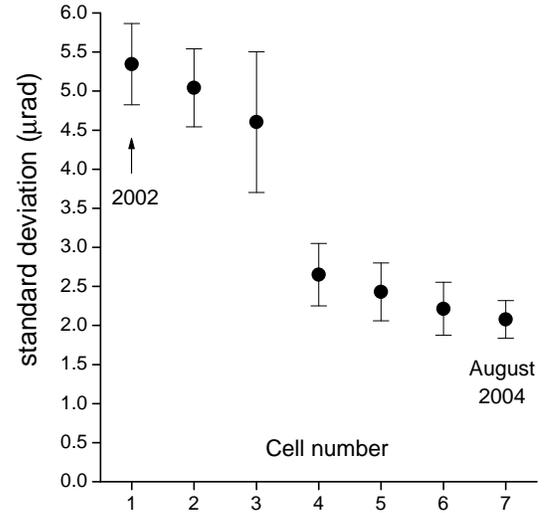}}
\caption{\label{noicel} Right: Standard deviation
SD of the distribution of   the experimental values  $\theta^{pv}_{exp}$($\mu$rad) obtained in each individual cell versus the
cell numbers chronologically ordered. The error bar on SD is estimated from the dispersion of the SD 's over their distribution in
one cell.         }
\end{figure}

%%%%%%%%%%%%
Figure~\ref{rescel} summarizes the experimental determinations of  $\theta^{pv}_{exp}$ obtained cell by cell in
seven different Cs cells, with their standard errors and the number of individual isotropic values $N_{iso}$ accumulated to obtain
each result. Figure~\ref{histo} presents the histogram for all the data obtained using the last four cells which have by
far the largest statistical weight. The detailed results obtained successively in the different cells are shown in Figure~\ref{detres}. Since
all the measurements were not performed at the same applied potential difference but most of them at a voltage $ 5\%$ lower, we have
made the appropriate correction for renormalizing all results at the same nominal value of
$E_l$, that of ref \cite{bou02}, 1619~V/cm. The SNR improvement from the first to the last cell is made conspicuous in Fig.~\ref{noicel}
which represents the standard deviation of the distribution of all data accumulated in each cell, versus the cell
number. Even so, this graph does not make apparent the additional factor
of improvement of the SNR per unit of time that results from an increase of the repetition rate.

It is important to test the agreement between
the results obtained with the seven different cells.  More precisely, we check whether the dispersion between the
means $m_k$ is compatible with the dispersion
$\sigma_k$ within the  measurements performed in each cell. To this end we form the quantity  $Q^2 =
\sum_k (  (m-m_k)^2 / \sigma_k^2 ) $ where 
$m = \sum_k (m_k/\sigma_k^2) /  (\sum_k 1/ \sigma_k^2  )$ is the weighted average.   $Q^2 $ is expected to be sampled from a
$\chi^2$ distribution with $\nu = K-1$ degrees of freedom, $K$ = 7 being the number of cells. 
We find $Q^2/\nu = 7.7/6 = 1.28$ (probability of exceeding 0.26).    
 Such an agreement suggests that possible defects, 
associated with the preparation of the cells (their geometry and surface properties \cite{noteC}, the filling
procedure, {\it etc}...)  have no detectable effect on our results. Indeed, all the cells were not made of
exactly the same material, sapphire/alumina, nor was the machining process identical. All cells had their
windows precisely mounted, normal to the tube except cell $\#$ 2 whose windows were tilted at
 $ + $ 3 and $ - $3 mrad towards the horizontal. This allowed us to align $\vec E_l$ precisely along the beam
direction (see Sect. III D 2) though this did not provide convincing advantages. Possible presence of foreign gas was
tested by looking for a broadening of the saturated absorption spectrum of the   
$D_2$ resonance line on an auxiliary set-up. Only in cell $\#$ 3 could it be observed. This might explain the presence of  
an unusual  short-term noise in this cell, which was rapidly discarded. Note, however, that the value of
$\theta^{pv}_{exp}$ from this cell still agrees with the average value.   

 
Our present result is:
\begin{equation}
 \theta^{pv}_{exp}(\mu{\rm rad}) = 0.950 \pm 0.025\, , ~~{\rm at}~~ E_l = 1.619 {\rm kV/cm}\, .   
\end{equation}
This value is nearly unaffected (relative difference 2$\times 10^{-3}$) if one attributes the same weight to each individual
datum instead of averaging the various runs made with each individual cell and then averaging the results in each cell over
the ensemble, with weights $\propto 1/\sigma^2$ at each stage. This gives us confidence that, at the quoted level of
precision, our measurements are unaffected by spurious properties varying from one data sample to another. 
Our data satisfy two other consistency tests:

i) agreement between the results obtained with 
$x$ and $y$  polarizations or $u$ and $v$, {\it i.e.} 
$S_{xy} = S_{uv}$ within the statistical uncertainty:
\begin{equation}
 \frac{1}{2}(S_{xy}^{av} - S_{uv}^{av}) =  0.006 \pm 0.025 \;\mu{\rm rad},    
\end{equation} 

ii) identity of the results of the two reconstitution methods using either $A_{LR}$ or $\vartheta$ (Eqs. 4 to 7).  

Within our uncertainty our result (Eq. 10) is in excellent agreement with the Boulder one \cite{woo97}, which predicts:
\begin{eqnarray}
\theta^{pv}(\mu{\rm rad}) = - \frac{{\rm{Im}} E_1^{pv}}{\beta E_l}= 0.962 \pm 0.005\, , \nonumber \\
~~{\rm at~~} E_l = 1.619 {\rm kV/cm}\, .   
\end{eqnarray}
for the hyperfine line $ 6S_{F=3} \rightarrow 7S_{F=4} $ explored during our measurements.

Combining our result, $ {\rm{Im}} E_1^{pv}/\beta  = - 1.538 \pm 0.040$ mV/cm,  with the value of the vector polarizability
determined in
\cite{ben99},  $\beta = 27.02 \pm 0.08 ~a_0^3$, we obtain:
\begin{equation}
{\rm{Im} }E_1^{pv}(6S_{F=3} - 7S_{F=4}) = - (0.808    \pm 0.021) \times 10^{-11} ~\vert e \vert a_0 \, .        
\end{equation}  
In addition to the statistical uncertainty, the quoted uncertainty includes the uncertainty in the estimation of the registered
systematic effects (Sect. IV) as well as in the determination of $E_l$ (Sect. V. D).
Thanks to our control of systematics and our gain of precision attained in the longitudinal field measurement, 
the absolute precision in $E_1^{pv}$ reached by our result is limited only by statistics and reaches $2\times 10^{-13}$ atomic
units. For comparison we note that an absolute precision of $3\times 10^{-12}$ atomic units was obtained by the most accurate
measurements of $E_1^{pv}$ performed in heavier atoms (Tl, Pb, Bi) where it is 30 times larger \cite{fla04}, but where
more difficult atomic physics calculations, presently less precise, are required to extract the weak charge.  
 
%%%%%%%%%%%%%% 
\section{Relevance of Atomic Parity Violation. Conclusion and prospects}
%%%%%%%%%%%%%%
\subsection{Goals for further APV measurements}

The main goal of atomic parity violation (APV) is to provide a determination of the weak nuclear charge $Q_W$, from the
measurement  of $E_1^{pv}$ {\it via} an atomic physics  calculation which now aims  
at 0.1$\%$ precision \cite{der04}. In view of present 
 and forthcoming results from high energy experiments, an important issue concerns the relevance of further
improving difficult experiments such as APV measurements. We would like to present arguments in favor of 
small scale APV experiments. 

  $ \bullet$~ 1 - 
 First  we wish to reiterate  that APV experiments explore the 
 electroweak (EW) electron-hadron  interaction  within  a range of low momentum transfers $q_{at}$   
 of 1 MeV or thereabouts in Cesium, which compares with the huge ones explored in collider experiments:
 100 GeV at LEP~I and LEP~II and 1 TeV at LHC \cite{notePP}.  At low
 energies, the electroweak amplitude is of the order  of $e^2 q_{at}^2/  M_W^2 $. 
In order to compensate  for this exceeding small factor, atomic  experiments have to be performed in very  special
conditions  (on a highly forbidden transition in a heavy atom). To obtain relevant  information, one has to approach 
an absolute precision of $10^{-8}$ in  the measurement of a radiative atomic transition  LR asymmetry.

  $ \bullet$~ 2 - 
 For $q_{at} \sim 1$ MeV, the quarks of the atomic nucleus act 
 {\it coherently}, while at high energies the nucleons are broken into their 
fundamental constituants: 
the quarks act then {\it incoherently}. This is what happens in deep inelastic 
electron-nucleon  scattering, such as the SLAC experiment \cite{pre78} involving a 
 GeV polarized electron beam colliding against a  fixed deuterium target. As a consequence, 
different  combinations of  electron-quark  PV  coupling constants are involved in the LR asymmetries
of  the two experiments:
  $\frac{2}{3}C_u^{(1)} -\frac{1}{3}C_d^{(1)}$ at high energies instead 
 of  $( 2Z + N) C_u^{(1)} + (Z+ 2N) C_d^{(1)} $ for $Q_W$.
 It is easily seen that, in a model-independent analysis, the two experiments delimit nearly orthogonal allowed bands in the
$\lbrack  C_u^{(1)},C_d^{(1)} \rbrack $ plane \cite{bou97}.

 $ \bullet$~ 3 - 
 Deviations $\Delta Q_W$ of $Q_W^{exp}$ from the SM prediction, are 
 most often analyzed in the framework of  {\it ``new physics''} models
 which affect EW interactions {\it at energies higher than $M_{Z_0} 
 c^2$} through the existence of gauge bosons heavier than the $Z_0$, such as for 
 instance Kaluza-Klein excitations of the SM  gauge bosons \cite{ant01}.  It turns out that  
 $\Delta Q_W$ is proportional to the same factor
 $X= \frac{\pi^2}{3} R_{\parallel}^2 M_{Z_0}^2$ as the deviations from
the SM  in  existing collider experiments, provided that $q^2 R_{\parallel}^2 \ll 1$, where
 $R_{\parallel} \leq 1$ TeV$^{-1}$  stands for the compactification 
 radius associated with the additional $d_{\parallel}$ dimensions 
 of the new physical space for EW gauge fields. 
A determination of
 $\Delta Q_W$  below the  0.1 ${\%} $ level of precision would give 
 constraints on $R_{\parallel}$, competitive with those of LEP II 
 \cite{all04,delg00,bou04a}. 
Furthermore, one can consider models which predict effects undetectable by LEP II results but that would be
visible in APV experiments \cite{delg00,che03}. Therefore, a 0.1$\%$ accurate $Q_W$ determination could allow
one  to impose a $\sim 5$ TeV limit to the compactification mass $R_{\parallel}^{-1}$  in a direction possibly invisible to
 high energy experiments.
 
 $ \bullet$~ 4 - 
 The fact that $q_{at}\sim $1 MeV allows one to investigate the 
 possible existence of \textit{ extra, neutral, light, gauge bosons }
  more precisely with a mass in the range  of a few MeV. Such a drastic  
modification of EW interactions  appears as  an alternative explanation  
for the remarkably intense  and narrow   gamma  ray line emitted  from the bulge of our galaxy, close to the energy  of 511
keV which coincides with the electron mass \cite{sil04,boe04}.  According to this
somewhat exotic  model, the observed spectrum 
 would result from the annihilation of two \textit{ light dark matter particles}
 (mass $\geq$ 1-2 MeV) into a pair $(e^{+}, e^{-})$ via the 
 exchange  of a light gauge boson U, with a mass 
 of about 10  MeV \cite{fay04}. 
 In order to reproduce the size of the effects observed experimentally, 
 one  has to  exclude at a large confidence level an
 {\it axial} coupling of the electrons to  the new  U boson, while such a coupling
is  the only possible   one for \textit{dark particles} which carry no charge.
This is  where APV comes into play.

The most plausible conclusion to which the present value of $\Delta Q_W$ leads \cite{fla04} is that 
the U boson couples  to the electron  as a vector particle with no axial coupling  at the $10^{-6}$ level,  
while its vector coupling to leptons and quarks are of the same  order of magnitude \cite{bou04b}. Thus, the APV 
measurements  provide  an empirical justification
 for  a key hypothesis, introduced  in  the astrophysical  model accounting  for the $ 511$ keV galactic line.

 \subsection{Conclusion and Prospects}
Our experiment has provided yet another method to measure atomic parity violation in a highly
forbidden transition. In the first Cs experiment \cite{bou82}, the signal detected was {\it the circularly polarized}
fluorescence intensity emitted on the $7S_{1/2}-6P_{1/2}$ transition. In an early version of their experiment
\cite{noe88}, the Boulder group detected the {\it total} fluorescence intensity emitted in the second step of the $7S-6P-6S$
cascade. In their final measurements \cite{woo97}, they operated with an atomic beam optically pumped in one hyperfine
state. They detected, by scattering of resonance photons, the population of the second hyperfine ground state resulting from
excitation of the forbidden transition followed by cascade deexcitation. However, this signal was superimposed on a
background ($\sim 25\%$) arising from stray resonant light. In all cases, the LR  asymmetry was finally observed {\it via   
fluorescence photons} and directly given by the ratio ${\rm Im} E_1^{pv}/ E_1^{Stark}$.      

 Our new method exploits the amplification
by stimulated emission of a resonant probe beam passing through the vapor along the path of the excitation beam
for the short time during which the 7S atoms have not yet decayed. The polarization of the probe is modified during
this propagation in a way which reveals the parity violating LR asymmetry, the key-point being that during the
propagation of the probe beam through the vapor the LR asymmetry itself is amplified exponentially. Consequently, the
measured asymmetry is no longer inversely proportional to the applied electric field, but rather an increasing function of it.
Moreover, the detected, differential, signal is directly the LR asymmetry with no background.   

During the course of the  work presented here, starting from the preliminary results
which validated the method
\cite{bou02}, we have succeeded in improving the SNR by a factor of 3.5. Our present result, still in 
agreement with the Boulder result, has now reached a relative accuracy on ${\rm Im} E_1^{pv}/
E_1^{Stark}$ of $2.6\%$, and an absolute precision of  2.5 $\times 10^{-8}$.   We have
described the main modifications of the apparatus that contributed to this gain in sensitivity.  We have also
shown how we can maintain good control of the systematic effects:  by making frequent measurements of
the $B_z$ field odd under
$\vec E_l$ reversal, and of the transverse $\vec E_t$ and $\vec B_t$ fields and by suppressing the effect of the tilt of the cell with respect to
the common beam axis. In addition, data analysis provides for confidence tests of  the results. Of particular relevance is the
compatibility of the results obtained in seven different cells which gives a rather good guarantee against systematic effects
arising from cell preparation, prone to variations from one cell to another. 

To interpret our data, we measured the electric field
experienced by the atoms inside the cell. To this end, we have performed the detection of the $7S$ state
alignment {\it in absence of any electric field} arising from the magnetic dipole contribution to the $6S-7S$ forbidden transition,
$\propto M_1^2$, an effect  unobserved heretofore. Since the detection of an alignment relies on the existence of 
hyperfine coupling in the two atomic states connected by the probe transition (the alignment signal cancels out without this
coupling), it is free of collisional background and molecular contribution and still more specific to the forbidden transition
than an orientation signal is. Therefore, it offers a nice way for extracting $E_l$ from the ratio  
$(\beta E_l/M_1)^2$ of the alignments measured with and without the field. 
 However, caution was needed to incorporate in the signal analysis
 existing deviations with respect to a pure exponential-type amplification process.~

~~~~~~~~~~~~          
 
We find it remarkable that results of APV experiments that involve scattering photons, of only 
a few eV, by a sample of a few cubic centimeters of dilute atomic vapor, can stand comparison with experiments
performed in colliders of the  highest energy, for providing a lower limit on the mass of a hypothetical additional
neutral boson.  In view of the present need for further measurements, underlined above, there are strong incentives 
to pursue APV measurements exploiting stimulated-emission detection:

$ \bullet$~1 - Given the difficult task of controlling and measuring systematic effects by
the Boulder group 
\cite{woo97, woo99}, a cross-check at the 1$\%$ level ({\it i.e.} 2 $\sigma$) of the 0.5$\%$ Boulder result for  
the $^{133}$Cs $ 6S_{F=3} \rightarrow 7S_{F=4}$ line by an independent method would constitute a valuable
result.  Such a statistical accuracy is now within reach with our set-up, even if no further SNR improvement were obtained.
Among all systematic effects registered so far, nothing indicates that they might have a redhibitory effect at the 1$\%$ precision
level, which therefore appears as an achievable goal. 
 
$\bullet$~ 2 - As shown in a recent paper \cite{bou05}, asymmetry amplification can provide a considerable enhancement factor in
a {\it transverse} field configuration and a longer interaction length. A cell with special multi-electrode design 
could "restore cylindrical symmetry", 
despite the application of a transverse $\vec E$ field. Then in a quantum noise limited measurement a  $0.1\%$
statistical precision would be achievable. Increasing further the probe optical gain would seem to be limited by
the onset of spontaneous superradiance, but {\it {triggered}} superradiance on the other hand would come into
play as a unique tool for even larger amplification of the asymmetry and possibly even better precision. 
The motivation for this project, which looks feasible, is encouraged by
the recent success of atomic theoretical physicists who were able to reduce their calculation uncertainty to the 0.5$\%$ level 
\cite{der01,mil01,fla02,fla04} in 2002, and by their considerable efforts now undertaken  
to arrive at 0.1$\%$ accuracy in their  many-body perturbation theory calculations \cite{der04}. 

$\bullet$~3 - In a cell experiment the required cesium quantity is very small, of order a few milligrams, 
 {\it i.e.} several orders of magnitude
smaller  than the required quantity in an effusive beam APV experiment \cite{woo99}. 
This opens the possibility of an APV measurement with
$^{135}$Cs, a radioactive isotope with a long half lifetime (3 million years). A quantity of 1~mg of $^{135}$Cs 
corresponds to an activity  of approximately 
$4\times 10^4$ Bq ($\approx 1~\mu$Ci), so that necessary radioprotection measures should not preclude the
feasibility of such an experiment. Measuring APV with two different isotopes, such as $^{135}$Cs and $^{133}$Cs  would
provide the very first experimental test of the nuclear weak charge dependence on the neutron number. Since the uncertainty
resulting from the neutron distribution is expected to be less than 0.1 $ \%$ in cesium \cite{der02}, the isotopic
dependence would offer an alternative interesting way of testing the Standard Model \cite{for}.
 
$\bullet$~4 - An independent measurement of the nuclear anapole moment, obtained from the
difference of the $E_1^{pv}$ determinations on  two different hyperfine $6S-7S$ lines today looks 
particularly necessary in view of the apparent inconsistency of the Boulder result \cite{CB91,fla04} with other data
relating to parity-violating nuclear forces. There exists a long-term project aiming at a direct measurement of the
nuclear anapole moment by searching for a linear Stark shift of alkali atoms trapped in a cristalline helium matrix
of hexagonal symmetry \cite{bou01}.  Even so, today our experiment has already reached a level of sensitivity such that
pursuing  this goal on the Cs $6S-7S$ transition appears achievable.       

\section*{Acknowledgements}

Laboratoire Kastler Brossel is a Unit\'e de Recherche de l'Ecole Normale Sup\'erieure et de
l'Universit\'e Pierre et Marie Curie, associ\'ee au CNRS (UMR 8552). F\'ed\'eration de recherche du
D\'epartement de Physique de l'Ecole Normale Sup\'erieure est associ\'ee au CNRS (FR 684).  

We gratefully acknowledge financial support from IN2P3 CNRS, without which this work would not have been 
possible. We are very grateful to S. Haroche, C. Cohen-Tannoudji, M. Voos, and M. Spiro for 
solving the administrative imbroglios  encountered by our group. We are indebted to M. Himbert, P. Juncar and
their collaborators of BNM-INM for financial help, generous loan of material and technical advice as well as to
V. Croquette from LPS-ENS for always available, efficient help. 

 It has been a pleasure to collaborate (through the support of
DRI/CNRS) with D. Sarkisyan and A. Papoyan, from the Research Institute of Ashtarak. Their efforts have been
decisive in improving the cells.   We also thank  S.~ Sanguinetti for participating with enthusiasm to the
modifications of the set-up at an early stage. M. A \"\i t Mohand provided help with infinite patience in the tedious
task of  numerical data analysis. We thank him as well as E. Brezin and F. Zomer for making this possible.    

We are most grateful to particle physicists M. Davier and J. Iliopoulos for their scientific interest and continuous
support.  Last but not least, most invaluable, has been our interaction with M.D. Plimmer.


\begin{thebibliography}{50}
\bibitem{notpul}
The $E_l$ field is pulsed to prevent the onset of discharges in the vapor.
\bibitem{note6P}
There are several non-resonant mechanisms for the direct excitation of a 6P population by the  excitation pulse. In the
present operating conditions (reduction of the dimer density thanks to a high body temperature of the cell, short
excitation-probe delay and large excitation energy $\sim$ 1.8 mJ), the dominant mechanism is
attributable to $6S-6P $ Frank-Hertz collisional  excitation by the photoelectrons emitted by the ``cathode'' window
and  accelerated by the applied longitudinal electric field.
\bibitem{noteM}
The magnetic dipole amplitude  $M^{'}_1$, includes two contributions: the larger one $M_1$ arises from many-body
and relativistic effects and the other one $M_1^{hf}$, induced by the non-diagonal hyperfine interaction, is present only in
$\Delta F = \pm 1$ transitions (note that the authors of \cite{ben99} use a sign convention for $M_1^{hf}$, opposite to our's
\cite{bou88}). Here the $E_2$ amplitude (see \cite{bou88}) does not contribute, since, in linear
excitation polarization with a quantification axis chosen along $\hat \epsilon_{ex} \wedge \hat k$ it can only induce $\Delta
m_F = \pm 2$ transitions, as opposed to the $M^{'}_1$ amplitude which induces only $\Delta m_F =0 $ transitions, hence $E_2$
and $M^{'}_1$  cannot interfere. Moreover, we can neglect $\vert E_2 \vert ^2/ \vert M_1 \vert ^2 \ll 10^{-3}$.
Note also the absence of any interference term,
$\propto \beta E_l  M^{'}_1$ in the transition probability, arising from the $\pi/2 $ phase difference between the Stark and
$M^{'}_1$ amplitudes.
\bibitem{noteC}
Before construction of the cells, the sapphire windows were controlled and rejected in case of  birefringence or
parallelism defects, then they were annealed at 1200$^{\circ}$C, a temperature at which we have checked that
reconstruction of the surface takes place. For cell $\#$ 4 and subsequent ones, the tube 
(either sapphire or alumina ceramics) was also annealed at 1450$^{\circ}$C.
\bibitem{notePP}
In this section we use the system of units commonly employed by particle physicists with $\hbar = c = 1$, according
to which a length is given by the inverse of a mass or a momentum.
\end{thebibliography}
\end{document}